\documentclass[12pt,preprint]{aastex}

\usepackage{amssymb}
\usepackage{graphicx}

\begin{document}

\title{The mid-IR Luminosity Function at z$<$0.3 from 5MUSES: Understanding the Star-formation/AGN Balance from a Spectroscopic View}
\author{Yanling Wu\altaffilmark{1}, Yong Shi\altaffilmark{1}, George Helou\altaffilmark{1}, Lee Armus\altaffilmark{2}, Daniel A. Dale\altaffilmark{3}, Casey Papovich\altaffilmark{4}, Nurur Rahman\altaffilmark{5}, Kalliopi Dasyra\altaffilmark{6}, Sabrina Stierwalt\altaffilmark{2}}
\altaffiltext{1}{Infrared Processing and Analysis Center, California
  Institute of Technology, 1200 E. California Blvd, Pasadena, CA
  91125} 
\altaffiltext{2}{Spitzer Science Center, California Institute of
  Technology, 1200 E California Blvd, Pasadena, CA, 91125}
\altaffiltext{3}{Department of Physics \& Astronomy, University of Wyoming, USA}
\altaffiltext{4}{George P. and Cynthia Woods Mitchell Institute for
  Fundamental Physics and Astronomy, Department of Physics and
  Astronomy, Texas A\&M University, College Station, TX 77843-4242 }
\altaffiltext{5}{Department of Astronomy, University of Maryland,
  College Park, MD, 20742}
\altaffiltext{6}{Irfu/Service d' Astrophysique, CEA Saclay, France}

\email{yanling@ipac.caltech.edu, yong@ipac.caltech.edu,
  gxh@ipac.caltech.edu, lee@ipac.caltech.edu, ddale@uwyo.edu,
  papovich@physics.tamu.edu, nurur@astro.umd.edu,
  kalliopi.dasyra@obspm.fr, sabrina@ipac.caltech.edu }

\begin{abstract}
We present rest-frame 15 and 24\,$\mu$m luminosity functions and the
corresponding star-forming luminosity functions at z$<$0.3 derived
from the 5MUSES sample. Spectroscopic redshifts have been obtained for
$\sim$98\% of the objects and the median redshift is $\sim$0.12. The
5-35\,$\mu$m IRS spectra allow us to estimate accurately the
luminosities and build the luminosity functions. Using a combination
of starburst and quasar templates, we quantify the star-formation and
AGN contributions in the mid-IR SED. We then compute the
star-formation luminosity functions at 15\,$\mu$m and 24\,$\mu$m, and
compare with the total 15\,$\mu$m and 24\,$\mu$m luminosity
functions. When we remove the contribution of AGN, the bright end of
the luminosity function exhibits a strong decline, consistent with the
exponential cutoff of a Schechter function.  Integrating the
differential luminosity function, we find that the fractional
contribution by star formation to the energy density is 58\% at
15\,$\mu$m and 78\% at 24\,$\mu$m, while it goes up to $\sim$86\% when
we extrapolate our mid-IR results to the total IR luminosity
density. We confirm that the active galactic nuclei play more
important roles energetically at high luminosities. Finally, we compare
our results with work at z$\sim$0.7 and confirm that evolution on both
luminosity and density is required to explain the difference in the
LFs at different redshifts.

\end{abstract}


\section{Introduction}
The unprecedented sensitivity of the {\em Spitzer} Space Telescope
\citep{Werner04} has opened a new window to explore the infrared (IR)
universe. IR luminous galaxies, discovered from ground-based
\citep{Rieke72} and space \citep{Soifer87} observations, constitute an
important population. While UltraLuminous InfraRed Galaxies (ULIRGs,
L$_{\rm IR}>10^{12}L_\odot$) only account for $\sim$5\% of the IR
luminosity in the local universe, their contribution becomes
increasingly important at higher redshift, e.g. Luminous InfraRed
Galaxies (LIRGs, $10^{11}L_\odot<L_{\rm IR}<10^{12}L_\odot$) are
responsible for 70\%$\pm$15\% of the energy density at z$\sim$1
\citep{Lefloch05} and ULIRGs become more important than LIRGs at
z$\sim$2.

Recent surveys with the Multiband Imaging Photometer for Spitzer
(MIPS) \citep{Rieke04}, as well as earlier observations with the {\em
  Infrared Space Observatory} (ISO), probe the dust emission at much
fainter levels than that has been reached by the {\em Infrared
  Astronomical Satellite} (IRAS). This has vastly improved our
understanding of galaxy evolution. Surveys with {\em IRAS} first
established the local benchmark for mid- and far-IR luminosity
functions (LFs) \citep{Soifer87, Saunders90, Rush93, Shupe98,
  Sanders03}. In the nineties, 15\,$\mu$m ISOCAM observations were
used to derive the 15\,$\mu$m LF \citep{Xu00} at low redshift, as well
as to study the evolution effects from the number counts
\citep{Elbaz99,Chary01}. The recent work of \citet{Bothwell11} has
constrained the slope of the IR and UV LFs at the extreme faint end
for the first time using large datasets of local galaxies, and has
derived the distribution function of star formation rate in the local
universe. Deep MIPS surveys carried out in the past few years
fertilized the ground for understanding the evolution of
LF. \citet{Lefloch05} illustrated the variation of 15\,$\mu$m LF in
the range of 0.3$<z<$1.2 and suggested that the comoving IR energy
density evolves dramatically, increasing with look-back time as
(1+z)$^{3.9\pm0.4}$ up to z$\sim$1
\citep{Caputi07,Reddy08,Magnelli09}.

IR bright galaxies emit the bulk of their energy as dust-reprocessed
light generated by dusty star formation (SF) or accretion onto the
supermassive blackholes, referred to hereafter as active galactic
nuclei (AGN). Obtaining information on the relative contribution of
SF/AGN is critical for understanding a galaxy's integrated
emission. \citet{Lefloch05} explored the star-formation history at
0.3$<$z$<$1.2. However, the MIPS 24\,$\mu$m photometry they used for
that study did not allow them to account for the AGN contribution, or
constrain the relationship between the stellar mass growth and
blackhole mass growth. Recent work of \citet{Magnelli09, Rujopakarn10,
  Fu10, Goto11} have identified AGN dominated sources and excluded
these objects from their samples to derive the SF LFs.  The shape of
LF depends on the rest-frame wavelength. UV/optical LF usually follows
the profile of a Schechter function \citep{Arnouts05, Ilbert05}, while
at mid/far-IR wavelengths, the bright end slope is observed to be
shallower than the exponential cut-off of a Schechter function
\citep{Soifer87, Rush93, Sanders03}. This is rather intriguing because
both UV and far-IR emission traces the same star-formation, and the
different shapes for the LFs are suggested to be a result of dust
extinction effect. The recent work of \citet{Fu10} claims that the
shallower slope of the IR LF could be due to the contribution of AGN
at the high luminosity end and when the AGN component is removed, the
SF LF can be fit with a Schechter function. This further motivates our
work of separating the SF and AGN emission in our objects to
understand their contribution to the LF.

To quantify the relative contribution of star-formation and AGN to the
infrared luminosities, spectral decompositions have been performed by
several groups \citep{Sajina08, Pope08, Murphy09,
  Menendez-Delmestre09}. In high mid-IR luminosity systems at
z$\sim$2, \citet{Sajina08} found an average AGN fraction of
$\sim20\%-30\%$ of total IR luminosity for strong PAH sources while
this number goes up to $\ge$70\% for weak-PAH sources. The study of
\citet{Pope08} of a sample of submm galaxies at similar redshift has
revealed at most a 30\% contribution from AGN at mid-IR
(5-11.5\,$\mu$m rest-frame). In the local universe, the contribution
of nuclear activity to the bolometric luminosity of ULIRGs has been
quantified with six independent methods by \citet{Veilleux09},
reaching an average AGN contribution of $\sim35\%-40\%$, whereas
\citet{Nardini08} suggest that intense star formation accounts for
85\% of the IR emission in local ULIRGs, with AGN contributing
15\%. It is clear that the relative contribution of SF/AGN varies in
galaxies of different luminosities \citep{Yuan10, Hopkins10}, with the
AGN playing more critical roles in more luminous systems.  It must
also be a function of wavelength, since the spectral energy
distribution (SED) of the two components are quite distinct. The
different selection criteria and decomposition techniques among
different samples and authors also add to the differences derived in
the relative contribution of SF and AGN.

Current studies on IR LFs are mostly based on MIPS 24\,$\mu$m
observations, which implies a heavy reliance on the SED library used
to make k-corrections and derive the monochromatic continuum or total
infrared luminosities. Recently, \citet{Fu10} have derived 8\,$\mu$m
and 15\,$\mu$m LFs, as well as the corresponding star-formation LF,
for a sample of z$\sim$0.7 objects, taking advantage of their IRS
spectra. The 5 Millijansky Unbiased Extragalactic Survey (5MUSES), an
infrared spectroscopic survey of 330 objects selected by their MIPS
24\,$\mu$m flux densities, provides an important sample for
understanding the infrared galaxy population (Helou et al. 2011, in
preparation). Although IR luminosity functions have been extensively
studied at low redshift, we find 5MUSES to be a unique sample for
deriving LF for the following reasons: 1. Being a mid-IR flux-limited
sample, we reach a wide range of infrared luminosity
($\sim$10$^9$L$_\odot$ to $\sim$10$^{12}$L$_\odot$). 2. We reach to
higher redshifts than previous samples, e.g. the RBGS
\citep{Sanders03}. 3. The IRS spectra of our sample and far-IR
measurements for most of them allow us to minimize the uncertainties
on k-correction and luminosity estimation associated with adopting
specific SEDs to be applied to all sources. 4. Last but not least, the
IRS spectra allow for a careful decomposition of every single source
into a SF and AGN component and estimation of their contribution to
the luminosity density. This will facilitate further studies on how
the SF/AGN fraction in the integrated luminosity density evolves in a
cosmological context.

The paper is organized as follows. In Section 2, we briefly describe
the sample selection, and then the infrared and optical data used in
this study. We introduce our methodology of using the 1/V$_{\rm max}$
method to derive luminosity function in section 3, where we also
demonstrate how we correct for the incompleteness of the 5MUSES
sample. In the same section, we discuss in detail our methods of the
SF/AGN decomposition of the IRS spectra and how we estimate the SF
contribution in a statistical sense.  The 15 and 24\,$\mu$m LFs, as
well as the corresponding SF LFs are presented in Section 4, together
with a discussion of how the star-formation fraction varies with
luminosity and wavelength. We summarize our conclusions in Section
5. Throughout this paper, we assume a $\Lambda$CDM cosmology with
$H_0=$70\,km\,s$^{-1}$, $\Omega_m$=0.27 and $\Omega_\lambda$=0.73.

\section{Observation and Data Reduction}
\subsection{The Sample}
5MUSES is a mid-infrared spectroscopic survey of 330 galaxies with
24\,$\mu$m flux densities 5\,mJy$<$f$_{\nu}$(24$\mu$m)$<$100\,mJy.
The sources are selected from the SWIRE (Elais-N1, Elais-N2, Lockman
Hole and XMM) and the Extragalactic First Look Survey (XFLS) fields,
covering a total area of 40.6 square degrees on the sky. It provides a
representative sample at intermediate redshift ($\langle z
\rangle=0.144$), previously unexplored by {\em Spitzer} since most of
the spectroscopic work was focused on nearby spiral galaxies (SINGS)
\citep{Kennicutt03}, local LIRGs and ULIRGs \citep{Armus07, Armus09,
  Veilleux09}, and the much fainter and more distant (z$\sim$2)
galaxies \citep{Houck05, Yan07}. A total of 1111 objects have
f$_{\nu}$(24$\mu$m) between 5 and 100\,mJy from the five survey fields
of 5MUSES excluding stars. In order to efficiently observe the objects
using the staring mode of IRS and include the largest fraction of a
galaxy's integrated light, only objects unresolved within an aperture
of d=10.5$\arcsec$ (corresponding approximately to the slit width of
the Long-Low module of IRS) are included in the final pool and this
results in a total of 800 sources. Then 330 objects are randomly
selected from the 800 final candidates.  The details of the sample can
be found in Helou et al. (2011, in preparation).

\subsection{Data Reduction}
We have obtained low-resolution mid-IR spectra for all 330 objects in
5MUSES using the Infrared Spectrograph on board the {\em Spitzer}
Space Telescope. Both Short-Low (SL: 5.2-14.5\,$\mu$m) and Long-Low
(LL: 14-37\,$\mu$m) modules are used, with spectral resolution of
64-128. The integration time on each object ranges from 300 to 960
seconds to achieve roughly the same SNR \citep{Wu10}.  The IRS data
are processed by the {\em Spitzer} Science Center data reduction
pipeline version S17 and our data reduction starts from the pipeline
products designated as ``basic calibrated data (bcd)''. The
two-dimensional spectrograms are median combined and then the
off-source sky regions are subtracted. After removing the sky
background, the spectrograms are cleaned with the IRSCLEAN package to
remove bad pixels and apply rogue pixel correction. Then the
background-subtracted cleaned spectrograms are reduced with the
optimal extraction method of the {\em Spitzer} IRS Custom Extractor
(SPICE) software to extract the 1-D spectra. The details on the
reduction of the IRS data can be found in \citet{Wu10}.

The IRS spectra allow us to derive redshifts for sources with mid-IR
emission and/or absorption features. We have also searched for optical
spectra of our sample in the literature.  A total of 50 5MUSES objects
either do not have redshift information from the literature, or their
IRS spectra are characterized by featureless power law continua, which
cannot yield redshifts. We obtained redshifts for some of these
sources by using the double spectrograph instrument on the Palomar 200
inch Telescope in 2009B and 2010A terms. The optical data are reduced
with the IRAF software using the standard routines for bias
subtraction, flat-fielding, sky background removal and wavelength
calibration. Finally, for sources with multiple emission line
features, redshifts are derived using all available emission lines.

\subsection{The Redshift Completeness}

\begin{figure}
  \epsscale{1.0}
  \plotone{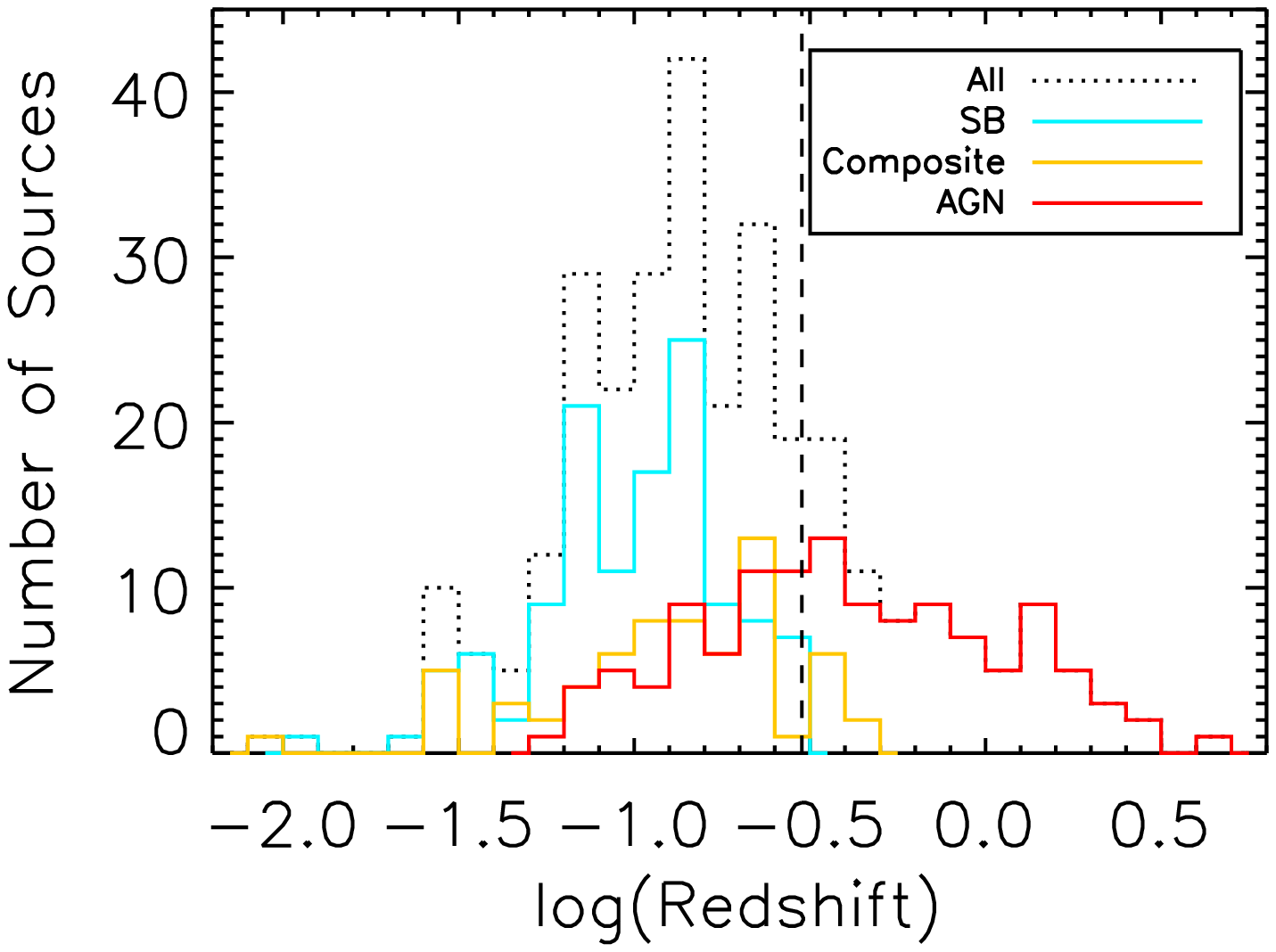}
  \caption{The Redshift distribution for 309 out of 330 sources in the
    5MUSES sample (dotted line). The blue, yellow and red lines
    represent the distribution for SB, composite and AGN-dominated
    sources. The dashed vertical line indicates the redshift cut of
    z$<$0.3 on which this paper is focused.}
  \label{fig:z_hist}
\end{figure}

\begin{figure}
  \epsscale{1}
  \plotone{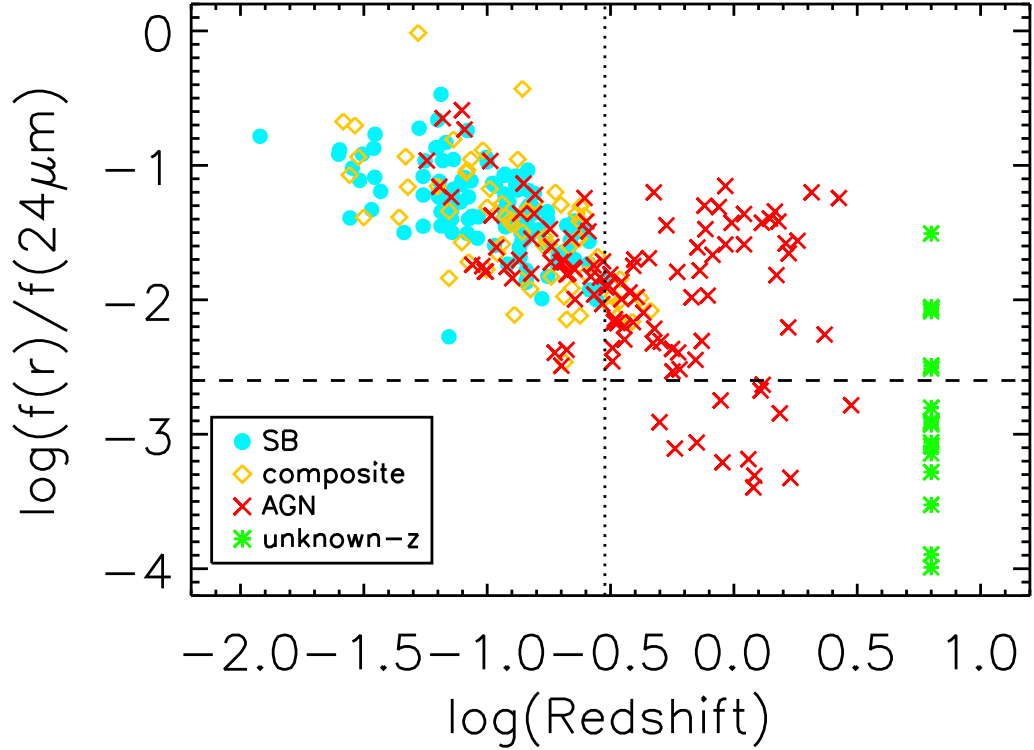}
  \caption{The ratio of flux densities at SDSS r band to the MIPS
    24\,$\mu$m band versus redshift for 5MUSES sources. The blue
    circles, yellow diamonds and red crosses represent the SB,
    composite and AGN sources. The green stars represent sources
    without redshifts. The dotted vertical line indicates our redshift
    cut of z$<$0.3. The dashed line indicates
    log[f$_\nu$(r)/f$_\nu$(24$\mu$m)]=-2.6. All 13 objects with
    log[f$_\nu$(r)/f$_\nu$(24$\mu$m)]$<$-2.6 are located at
    z$>$0.3. 17 out of 21 unknown-z sources have
    log[f$_\nu$(r)/f$_\nu$(24$\mu$m)]$<$-2.6, and are thus most likely
    to lie at z$>$0.3.}
  \label{fig:z_rz}
\end{figure}

Combining the IRS, Palomar spectra and literature work, we found
secure redshift measurements for 309 objects out of the 330 sources in
5MUSES (94\%). The redshift distribution of 5MUSES is shown in Figure
\ref{fig:z_hist}. We also show the redshift distribution for
starburst, composite and AGN-dominated sources respectively, which
have been classified based on their apparent 6.2\,$\mu$m PAH
EWs\footnote{Sources with 6.2\,$\mu$m PAH EW$>$0.5\,$\mu$m are
  SB-dominated; sources with 0.2\,$\mu$m$<$PAH EW$<$0.5\,$\mu$m are
  composite; and sources with PAH EW$<$0.2\,$\mu$m are AGN-dominated.}
(Wu et al. 2010).  Since we are only interested in the redshift range
of z$<$0.3 for this study, the relevant redshift completeness is close
to 1 for two reasons: (1) The sources for which we are not able to
find redshifts from 80 minutes of integration time on the Palomar 200
inch telescope have very low r band to 24\,$\mu$m band flux ratios
f$_\nu$(r)/f$_\nu$(24$\mu$m). As can be seen from Figure
\ref{fig:z_rz}, objects with low f$_\nu$(r)/f$_\nu$(24$\mu$m) ratios
(ie. log[f$_\nu$(r)/f$_\nu$(24$\mu$m)]$<$-2.6) are more likely to have
high redshifts (z$>$0.3) \citep [see also][]{Dey08}; (2) The IRS
spectra of the sources without redshifts are characterized by
featureless power-law continuum in the mid-IR. This indicates that
they are most likely AGN-dominated. It can be seen from Figure
\ref{fig:z_hist} that the median redshift for AGN-dominated sources is
much higher than SB or composite sources.  The median redshift for
sources with 6.2\,$\mu$m PAH EW$>$ 0.2\,$\mu$m is 0.13, while it is
0.40 for sources with 6.2\,$\mu$m PAH EW $<$ 0.2\,$\mu$m. Thus these
power law sources are much more likely to lie at the high end of the
redshift distribution.

For this study, we focus on objects with z$<$0.3, which includes 226
objects. Among the 21 (330-309) sources without redshift, only 4 do
not have very low f$_\nu$(r)/f$_\nu$(24$\mu$m) ratios, e.g. their
log[f$_\nu$(r)/f$_\nu$(24$\mu$m)]$>$-2.6, and might be located at
z$<$0.3. This indicates that the redshift completeness for our sample
at z$<$0.3 is $>\sim$98\%.

\section{Methodology}

\subsection{The Incompleteness Correction}
The targets for 5MUSES are randomly selected based on
f$_\nu$(24$\mu$m)$>$5\,mJy after excluding the resolved
objects. Understanding the selection function for 5MUSES is crucial
for building the luminosity function. Because we exclude extended
objects, it is likely that we have excluded more nearby objects than
those at higher redshifts. Thus when we derive the number density,
instead of applying a uniform correction factor of $\sim$3.4
($1111/330$), we need to investigate the selection effect in
individual redshift bins before we build our luminosity function.

Redshift information is not available for all the sources in the
parent sample of 5MUSES, thus we use the redshift catalog of
\citet{Papovich06} for the XFLS field to understand the redshift
distribution when a flux limit of 5\,mJy is
imposed. \citet{Papovich06} observed the XFLS field using the
Hectospec instrument on MMT for 5 positions covering a 1 degree
diameter field of view individually. Then they combined redshifts from
Hectospec with redshifts from SDSS, and reached a completeness of
$\sim$90\% at f$_{\nu}$(24$\mu$m)$>$1\,mJy in the 3.3 square degrees
of their survey field. These authors also provided the completeness
factors at different flux limits, which we use to derive the final
number counts at f$_{\nu}$(24$\mu$m)$>$5\,mJy in the XFLS field. Then
we divide the number of objects in different redshift bins by the
total number of objects in the XFLS field, and derive the fractional
contribution of number counts in this field at
f$_\nu$(24$\mu$m)$>$5\,mJy. Using this as a reference, we predict the
number of objects in the corresponding redshift bins for the 5MUSES
sample. Then we divide the predicted number counts by the number of
objects we have observed and obtain the correction factor in each
redshift bin. Finally, we fit a second-order polynomial to the data
and this gives us the completeness correction factor $\omega_i$(z),
which is then used to correct for the incompleteness at different
redshifts.


\subsection{The 1/V$_{\rm max}$ Method}
We limit our study of the mid-IR luminosity function to z$<$0.3
because the rest-frame 24\,$\mu$m band\footnote{Here we refer to the
  MIPS 24\,$\mu$m filter, instead of the monochromatic 24\,$\mu$m
  continuum.} moves outside the IRS spectrum beyond z$=$0.3.  In
addition to that, our relatively bright flux limit of
f$_\nu$(24$\mu$m)$>$5\,mJy results in a fast decrease of the number of
objects as redshift increases, which would yield results that have low
statistical significance at high redshift.

In this study, we use the 1/V$_{\rm max}$ method \citep{Schmidt68} to
derive the luminosity function, which does not require any assumption
on the shape of the LF. The 1/V$_{\rm max}$ method counts galaxies
within a volume. V$_{\rm max}$ is calculated individually for each
source in our sample as the maximal volume within which that galaxy is
detectable in this survey. The availability of 5-35\,$\mu$m IRS
spectrum allows us to accurately make k-corrections based on the
observed SED shape for individual galaxy. We first derive the
k-correction for each object, and then move the galaxy to the redshift
where its 24\,$\mu$m observed flux reaches the limit of this sample,
5\,mJy. The maximum comoving volume is calculated as:

\begin{equation}
  {\mathrm{V_{i,max}}}=\int_{\rm z_{low}}^{\rm z_{high}}\frac{dV}{dz}{dz}
\end{equation}
where [z$_{\rm low}$,z$_{high}$] is the redshift range of
interest. For our study, we ignore sources with z$<$0.02, while
z$_{\rm high}$ is the lower of the two: a) the maximum redshift
considered in this study, 0.3, b) the maximum redshift a source can be
detected at the limiting observed 24\,$\mu$m flux of 5\,mJy. The
uniformity of the distribution of galaxies is tested by checking the
V/V$_{\rm max}$ values and we find $\langle V/V_{\rm max}
\rangle$=0.54 for the sources used in this study.

\begin{figure}
  \epsscale{1.0}
  \plotone{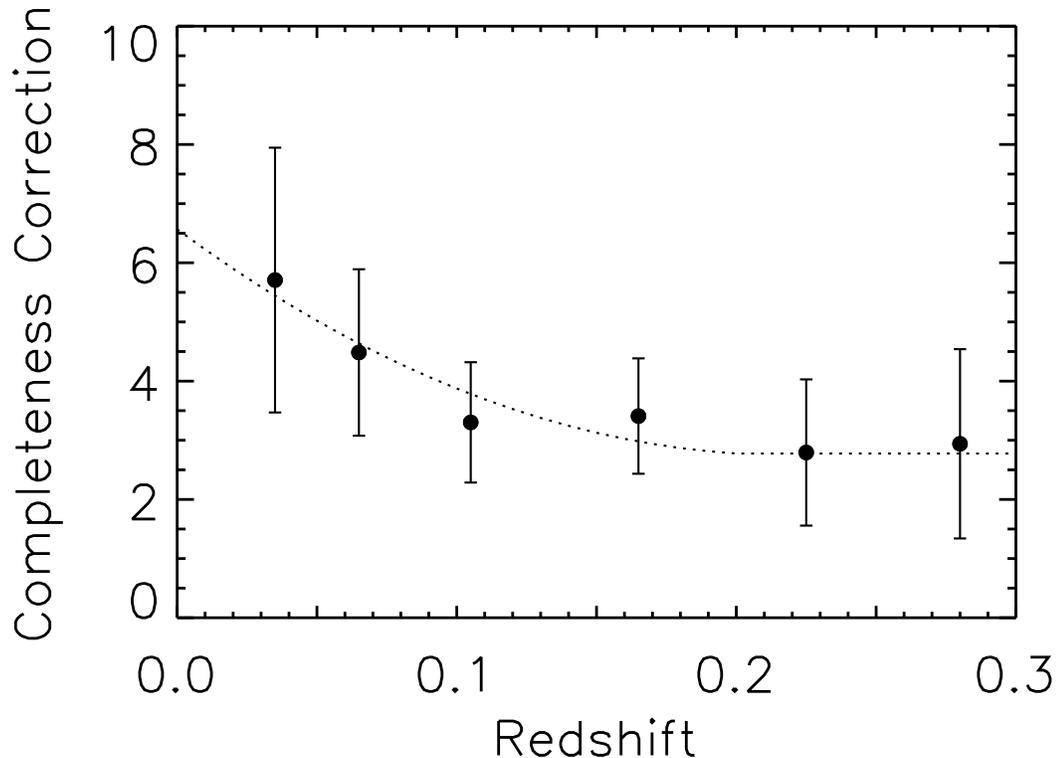}
  \caption{The completeness correction factor for 5MUSES versus
    redshift. Instead of using an average correction factor of
    $\sim$3.4 (1111/330), we benchmark our sample with the XFLS, where
    redshifts are available for this field. We first derive the
    fractional contribution of number counts in different redshift
    bins for the XLFS field at f$_{\nu}>$5\,mJy. Using this as a
    reference, we predict the number counts in each redshift bin for
    the 5MUSES sample. Then we divide the predicted number of objects
    by the observed number and obtain the completeness correction in
    each redshift bin. Finally, we use a second order polynomial to
    fit the data and obtain the completeness correction factor as a
    function of redshift. The error bars represent the Poisson noise.}
  \label{fig:selection}
\end{figure}

The luminosity function is then derived by using the following 
formula \citep{Schmidt68}:
\begin{equation}
  {\phi}=\frac{\mathrm{4}\pi}{\Omega}{\Sigma}\frac{\mathrm{1}}{\mathrm{V_{i,max}}}\frac{1}{\mathrm \Delta\mathrm{logL}}{\mathrm{\omega_{i}}}
\end{equation}
Where $\Omega$ is the total survey area of 5MUSES sample (40.6 square
degrees), V$_{\rm i,max}$ is the comoving volume over which the ith
galaxy could be observed, $\Delta$logL is the size of the luminosity
bin and $\omega_i$ is the completeness correction factor for the ith
galaxy. $\omega_i$ is a function of redshift and was calculated in
Section 3.1.  We divide the sources into seven luminosity bins, and
calculate the value of $\phi$ in each bin. The uncertainties include
both the Poisson noise statistics on the number of sources used in the
measurement, and the uncertainty associated with the completeness
correction factor $\omega_i$. As can be seen from Figure
\ref{fig:selection}, the uncertainty of $\omega_i$ is rather large,
mainly due to the small number of objects in each redshift bin in XFLS
at f$_\nu$(24$\mu$m)$>$5\,mJy, so we assign an uncertainty of 40\% to
$\omega_i$, which is the average uncertainty for the data points we
use to calculate the correction factor. Because our k-corrections are
made directly from the source SED, we have almost negligible
uncertainties associated with the conversion from the observed flux to
the rest-frame luminosities. For the luminosity functions presented in
this paper, we do not include noise from the cosmic variance since we
sample several widely distributed directions.


\subsection{Spectral decomposition in the mid-IR} 
\begin{figure}
  \epsscale{1.1}
   \plotone{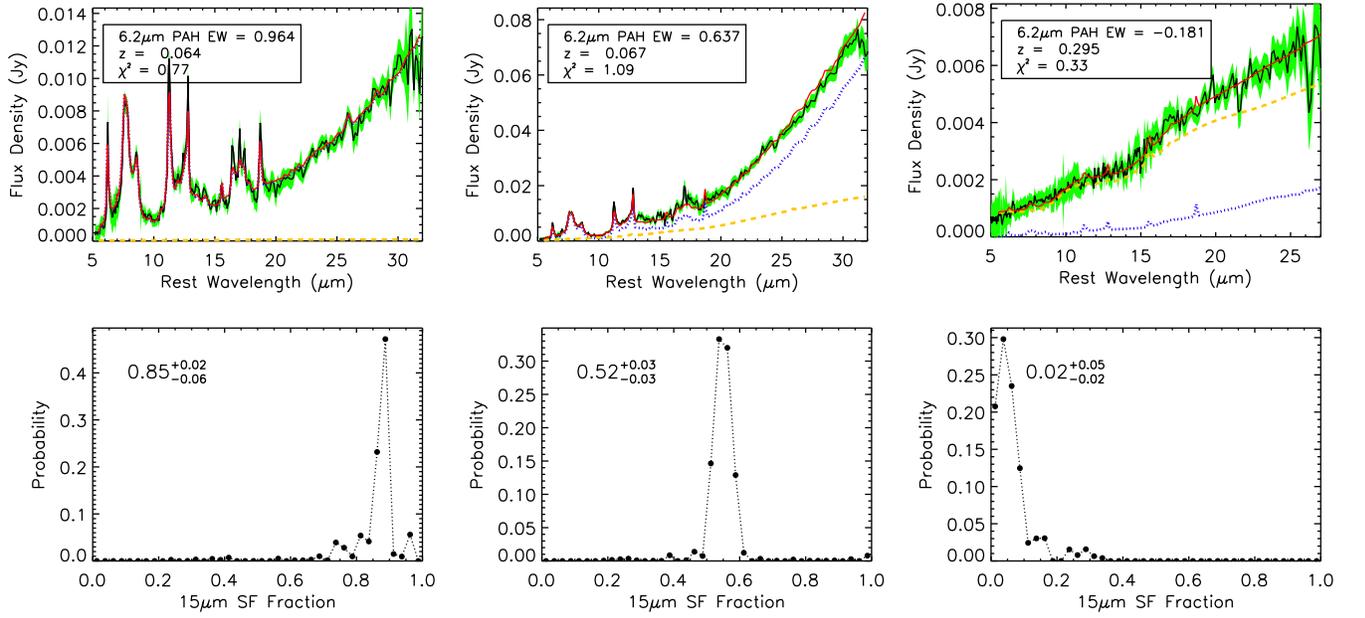}
  \caption{a) Upper panels: Examples of our decomposition on the IRS
    spectra for typical starburst, composite and AGN-dominated sources
    in 5MUSES. The black solid line is the IRS spectrum and the green
    shaded region indicates the uncertainties. The blue dotted line is
    the scaled star-formation galaxy template; the yellow dashed line
    is the scaled quasar template; the red solid line is the best-fit
    composite spectrum. The 6.2\,$\mu$m PAH EWs (a negative value
    indicates an upper limit), redshifts of the objects, as well as
    the reduced $\chi^2$ values are also shown on the plot. b) Lower
    panels: The probability density function for the star-formation
    fraction at 15\,$\mu$m for each corresponding object. The median
    fraction and its 16th to 84th percentile range is also indicated.}
  \label{fig:samplefit}
\end{figure}

The mid-IR SED of star forming galaxies and AGN show distinctly
different spectral features. As a result, the availability of
5-35\,$\mu$m IRS spectra for the 5MUSES sample allows us to
disentangle the star-formation and AGN contribution to the galaxy
luminosity. Star-forming galaxies often display broad emission
features, which are generally attributed to the emission from
PAHs. AGN, on the other hand, are usually characterized by featureless
power-law continuum (except for a few high-ionization fine-structure
lines) and their mid-to-far IR continuum slopes are normally flatter
than star-forming galaxies. A combination of one starburst template
and a power-law continuum with a free spectral index have often been
used to decompose galaxy spectra into star-formation and AGN
components for high-redshift galaxies \citep{Sajina08, Pope08,
  Menendez-Delmestre09, Murphy09}. The high S/N IRS data of the 5MUSES
sample allow us to take into account detailed mid-IR features and
decompose the galaxy spectra much more accurately. Among mid-IR
spectral features (PAH strength, continuum slopes, silicate strength,
fine-structure lines, etc.), PAH strength or the IR continuum slopes
are arguably the most commonly used indicators for star-formation, so
we select our templates mainly based on these two parameters. We start
from our empirical SED template library \citep{Wu10}, which has
included star-forming galaxies, ULIRGs and PG/2MASS quasars, and
select 15 star-forming galaxy (6.2\,$\mu$m PAH EW$>$0.5\,$\mu$m)
templates and 10 quasar templates, covering as large a range of PAH
strength and slope variation as possible. Then we perform a
least-$\chi^2$ fit for each combination of a star-formation template
and a quasar template to find the most likely coefficients that would
describe the observed 5MUSES spectrum as a linear combination of the
two templates. In the upper panels of Figure \ref{fig:samplefit}, we
show examples of decomposition of IRS spectra of typical starburst
dominated, composite and AGN dominated sources.

\begin{figure}
  \epsscale{1.1}
   \plotone{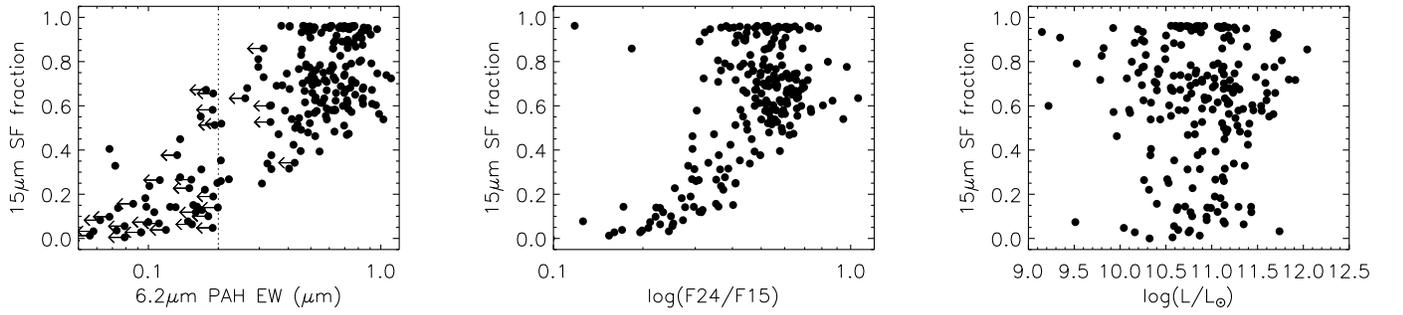}
  \caption{Left panel: The SF fraction at 15\,$\mu$m versus the
    6.2\,$\mu$m PAH EW. The dotted line indicates 6.2\,$\mu$m PAH
    EW=0.2\,$\mu$m.  Middle panel: The SF fraction versus the mid-IR
    continuum slope of f$_\nu$(24$\mu$m)/f$_\nu$(15$\mu$m). Both the
    PAH EW and continuum slope are correlated with the SF fraction at
    15\,$\mu$m with scatter. The two sources on the middle panel with
    very low f$_\nu$(24$\mu$m)/f$_\nu$(15$\mu$m) ratios and high SF
    fractions have their mid-IR spectra dominated by PAH emission,
    while their continuum slopes do not rise quickly. This also
    indicates that continuum slope alone has a high uncertainty when
    it is used as a SF indicator. Right panel: The SF fraction versus
    the infrared luminosity. There is no clear correlation between
    these two parameters. Note that at the high luminosity end, we
    don't observe many objects with low SF fraction. This is because
    we have not included very high luminosity sources in this study
    (z$<$0.3).}
  \label{fig:pahslopesf}
\end{figure}

{\bf Statistical constraints on the star-formation contribution: }
Rather than directly adopting the star-formation fraction at
15\,$\mu$m from the best-fit, we build the probability density
function ({\em pdf}) of the star formation fraction by weighting the
values of star formation fraction for each trial fit by
exp(-$\chi^2/2$). Then the star-formation fraction is taken to be the
median of the resulting probability density function and the 1$\sigma$
uncertainty is taken to be the 16th-84th percentile range. On the
lower panels of Figure \ref{fig:samplefit}, we show the corresponding
{\em pdf} for each source and the SF fraction and its associated
uncertainties are also indicated on the plot. As discussed earlier,
the 6.2\,$\mu$m PAH EW and the continuum slope have often been used as
indicators of star-formation activities. We compare the 6.2\,$\mu$m
PAH EW and continuum slope of f$_\nu$(24$\mu$m)/f$_\nu$(15$\mu$m)
versus SF fraction estimated from the probability distribution on the
left and middle panels of Figure \ref{fig:pahslopesf}. Our median
likelihood SF fraction clearly correlates with both parameters while
the scatter is quite significant for intermediate values of
f$_\nu$(24$\mu$m)/f$_\nu$(15$\mu$m) ratios. Note however that for
f$_\nu$(24$\mu$m)/f$_\nu$(15$\mu$m)$<$0.3, all but 2 sources have very
low star-formation fraction, suggesting that very flat mid-IR slopes
are a strong discriminator for AGN. We also note that some sources,
even though they have very large PAH EWs ($>$0.5\,$\mu$m), have
star-formation fractions only $>\sim$0.5. This could be due to the
fact that we are looking at the star-formation fraction at 15\,$\mu$m
in this study, while the contribution to the total IR luminosity is
very likely to be dominated by SF. On the other hand, we also need to
point out that the SF fraction we derive has at least $\sim$20\%
uncertainty. Finally, we plot the SF fraction versus the IR luminosity
for each source on the right panel of Figure \ref{fig:pahslopesf} and
we do not observe any correlation. This indicates luminosity itself
provides limited information on the energy source of a
galaxy. Clearly, there is no single parameter that could be used to
determine the SF fraction accurately, thus a combination of a few is
indeed needed in order to constrain the relative contributions of SF
and AGN.

\section{Results}
\subsection{The 24\,$\mu$m and 15\,$\mu$m Luminosity Functions at z$<$0.3}

\begin{figure}
  \epsscale{1} \plotone{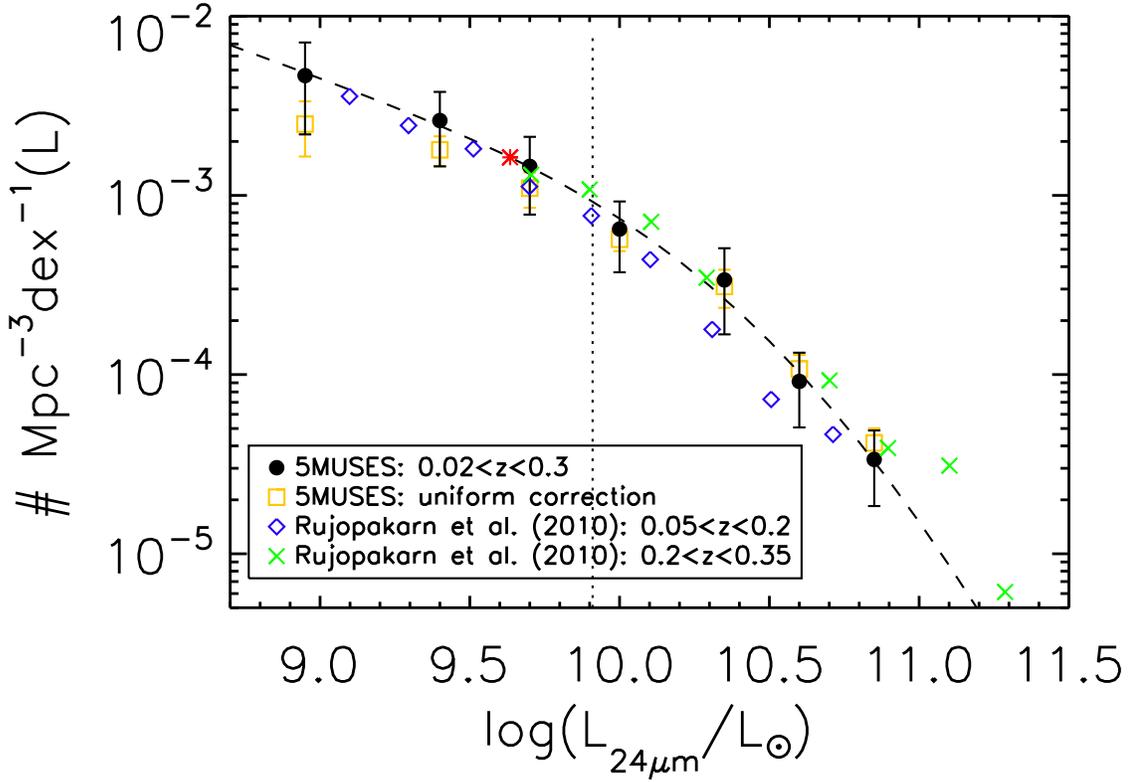}
  \caption{The rest-frame MIPS 24\,$\mu$m luminosity function at
    z$<$0.3 derived from the 5MUSES sample using the 1/V$_{\rm max}$
    method (black circles). The yellow squares represent the LF if we
    were to apply a uniform completeness correction factor of
    $\sim$3.4. The blue diamonds and green crosses represent the
    24\,$\mu$m LFs at 0.05$<$z$<$0.2 and 0.2$<$z$<$0.35 derived by
    \citet{Rujopakarn10}. The dashed line is the fit to the data
    points of our 24\,$\mu$m LF assuming a double power-law
    exponential function. The knee (L$^\star$) of the LF is
    represented by the red star on the plot. The dotted line denotes
    the luminosity calculated at the median redshift of this sample
    corresponding to the flux limit at 24\,$\mu$m. }
  \label{fig:LF24}
\end{figure}

From the 5-35\,$\mu$m IRS spectrum of the 5MUSES sample, we can
directly estimate the rest-frame MIPS-equivalent 24\,$\mu$m
luminosities. Using the 1/V$_{\rm max}$ method, we have derived the
24\,$\mu$m luminosity function for sources with z$<$0.3. Our
24\,$\mu$m LF is shown in Figure \ref{fig:LF24} and the corresponding
data points are reported in Table \ref{LFfit}.
We adopt a double power-law exponential function \citep{Saunders90}
to fit our LF:
\begin{equation}
\phi(L)=\frac{dN(L)}{dVdlog_{\mathrm{10}}(L)}=\phi^\star({\frac{L}{L^\star}})^{1-\alpha}exp{\{-\frac{1}{2\sigma^2}log_{\mathrm{10}}^2[1+(\frac{L}{L^\star})]\}}
\end{equation}

The dashed line is the fit to our data by running mpfit.pro. The
uncertainties on the fitting parameters are derived with 1000 Monte
Carlo realizations and they are reported in Table \ref{LFpara}. The
dotted line denotes the luminosity calculated at the median redshift
of our sample corresponding to the 24\,$\mu$m flux limit. Note that
5\,mJy is the flux limit on the observed 24\,$\mu$m, while the flux
limit we use here is taken to be the rest frame 24\,$\mu$m flux
density corresponding to the galaxy with f$_{\rm
  obs}$(24\,$\mu$m)=5\,mJy and maximum k-corrected. On Figure
\ref{fig:LF24}, we have also included the 24\,$\mu$m LFs by
\citet{Rujopakarn10} for comparison. Our LF is in good agreement with
\citet{Rujopakarn10} at the low end, while at the high end, our data
points are located between the 0.05$<$z$<$0.2 and the 0.2$<$z$<$0.35
LFs they have derived. This indicates that in the redshift range we
derive our LF, evolution is probably already at work, as will be
addressed in more detail later on.



\begin{figure}
  \epsscale{1}
  \plotone{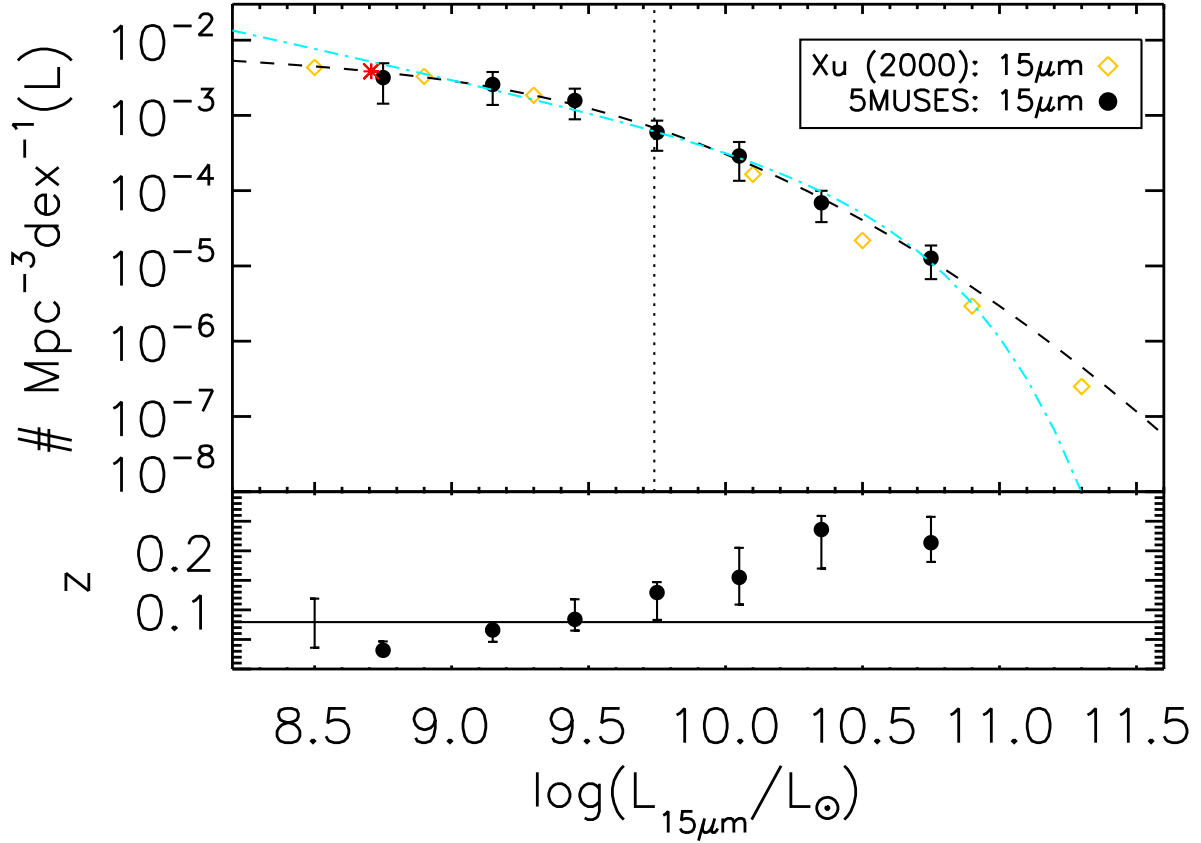}
  \caption{Top panel: The monochromatic 15\,$\mu$m luminosity function
    at z$<$0.3 derived from the 5MUSES sample (filled circles). The
    dashed line is the fit to the data points assuming double
    power-law exponential function. The blue dot-dashed line
    represents the fit with a Schechter function to our data. The knee
    (L$^\star$) of the LF is represented by the red star. The dotted
    line corresponds to the luminosity calculated at the median
    redshift of our sample for a galaxy at the flux limit. The local
    15\,$\mu$m luminosity function from \citet{Xu00} is overplotted
    with the yellow diamond for comparison. Our 15\,$\mu$m LF is in
    excellent agreement at the faint end with the ISO 15\,$\mu$m LF by
    \citet{Xu00}, but is slightly higher at the
    bright end. Bottom panel: The median redshift in each luminosity
    bin we have used to derive our LF. The solid line represents the
    median redshift of Xu's sample.}
  \label{fig:LF15}
\end{figure}

We then compute the rest-frame 15\,$\mu$m monochromatic\footnote{The
  monochromatic luminosity is calculated within a width of
  1\,$\mu$m.} luminosity function for our sample.
LF at this wavelength has been extensively studied in the local
universe \citep{Xu00, Pozzi04, Matute06} as well as at high redshift
\citep{Lefloch05, Magnelli09} to explore the evolution of galaxy
populations. We repeat the steps used for calculating the 24\,$\mu$m
LF and derive the 15\,$\mu$m LF for the 5MUSES sample. In the upper
panel of Figure \ref{fig:LF15}, we plot our 15\,$\mu$m LF with solid
circles and the corresponding data points are reported in Table
\ref{LFfit}. The dotted line denotes the luminosity calculated at the
median redshift of this sample corresponding to the 15\,$\mu$m flux
limit. As expected, the completeness level at 15\,$\mu$m is much lower
than at 24\,$\mu$m, since this is a 24\,$\mu$m selected sample at
z$<$0.3. The dashed line is the fit to our LF with double power-law
exponential function. During our fit, we fixed the faint end slope to
$\alpha$=1.2, which has been well determined from similar studies in
the local universe. The results of the fitting parameters are reported
in Table \ref{LFpara}. For comparison, we also overplot the local
15\,$\mu$m LF from \citet{Xu00} derived using ISOCAM
observations. Similar to what we have observed when we compare our
24\,$\mu$m LF with \citet{Rujopakarn10}, the 5MUSES LF is in good
agreement with Xu's LF at the low end, while our LF is slightly higher
at the high luminosity end. This can be explained by the differential
evolution effect. The density or luminosity evolution with redshift
have been extensively studied \citep{Lefloch05, perez-Gonzalez05,
  Magnelli09}. Evolution affects different luminosity bins by
different amounts, because they are populated from different redshift
ranges. In the lower panel of Figure \ref{fig:LF15}, we plot the
median redshift in each luminosity bin as a solid circle and the 16th
and 84th percentile as the error bar. The median redshift (z=0.09) for
Xu's sample has also been plotted (solid line) together with its
1$\sigma$ population dispersion. Although the median redshift for the
sources used in this study is 0.12, it is clear that the high
luminosity end is dominated by sources at higher redshift, thus our
slightly higher LF at the bright end is almost certainly a result of
that difference.

In addition to using the double power-law to fit the LF, we have also
attempted using a single Schechter function. In Figure \ref{fig:LF15},
we overplot the fit with Schechter function as the blue dot-dashed
line. Although Schechter function can fit our data reasonably well,
the fit is rather poor in both the faint and bright end when the data
points from the ISO 15\,$\mu$m LF by \citet{Xu00} are included.

\subsection{The Star-formation Luminosity Function}

In the previous subsection, we have shown that the 15\,$\mu$m LF need
to be fit with a double power-law exponential function because the
bright-end slope of the LF is clearly flatter than the Schechter
Function.  This is in contrast to UV LFs, which display much steeper
slopes at the high luminosity end. As both the UV and IR luminosities
trace active star formation, with the IR being the reprocessed portion
of the UV, one would expect similar behaviors by the LFs of UV and IR
emission. Using Spitzer IRAC observations of the Bootes field,
\citet{Huang07} have shown that the 8\,$\mu$m luminosity function of a
sample of {\em star-forming} galaxies does indeed follow the shape of
a Schechter function. More recently, using AKARI data, \citet{Goto11}
have shown that after removing the optically identified AGN, their IR
LF becomes much steeper. Using {\em Spitzer} IRS spectra, \citet{Fu10}
have studied the 15\,$\mu$m LF at z=0.7 and constrained the slope of
the SF LF at the high luminosity end using a Schechter function. So
the flatter LFs we are deriving probably reflect AGN contributions to
the IR, and we will verify this hypothesis in what follows. In Section
3.3, we have shown our method of decomposing the star-formation/AGN
contribution from the mid-IR spectra of 5MUSES sources and estimated
the star-formation fraction by taking the median likelihood of the
probability density function. Now we derive the star-formation
luminosity functions at 15\,$\mu$m and 24\,$\mu$m for 5MUSES at
z$<$0.3.

\begin{figure}
  \epsscale{1.2}
  \plottwo{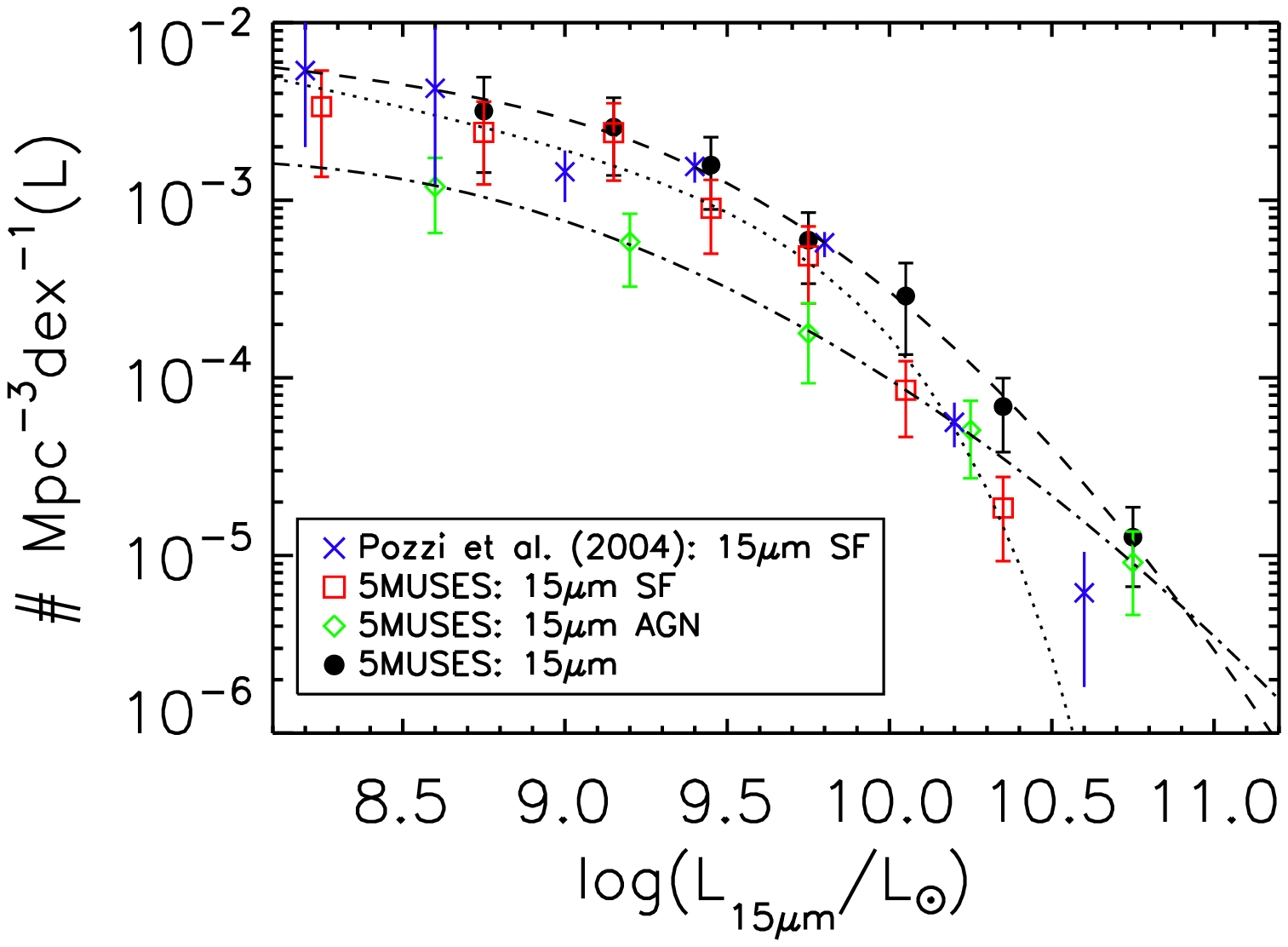}{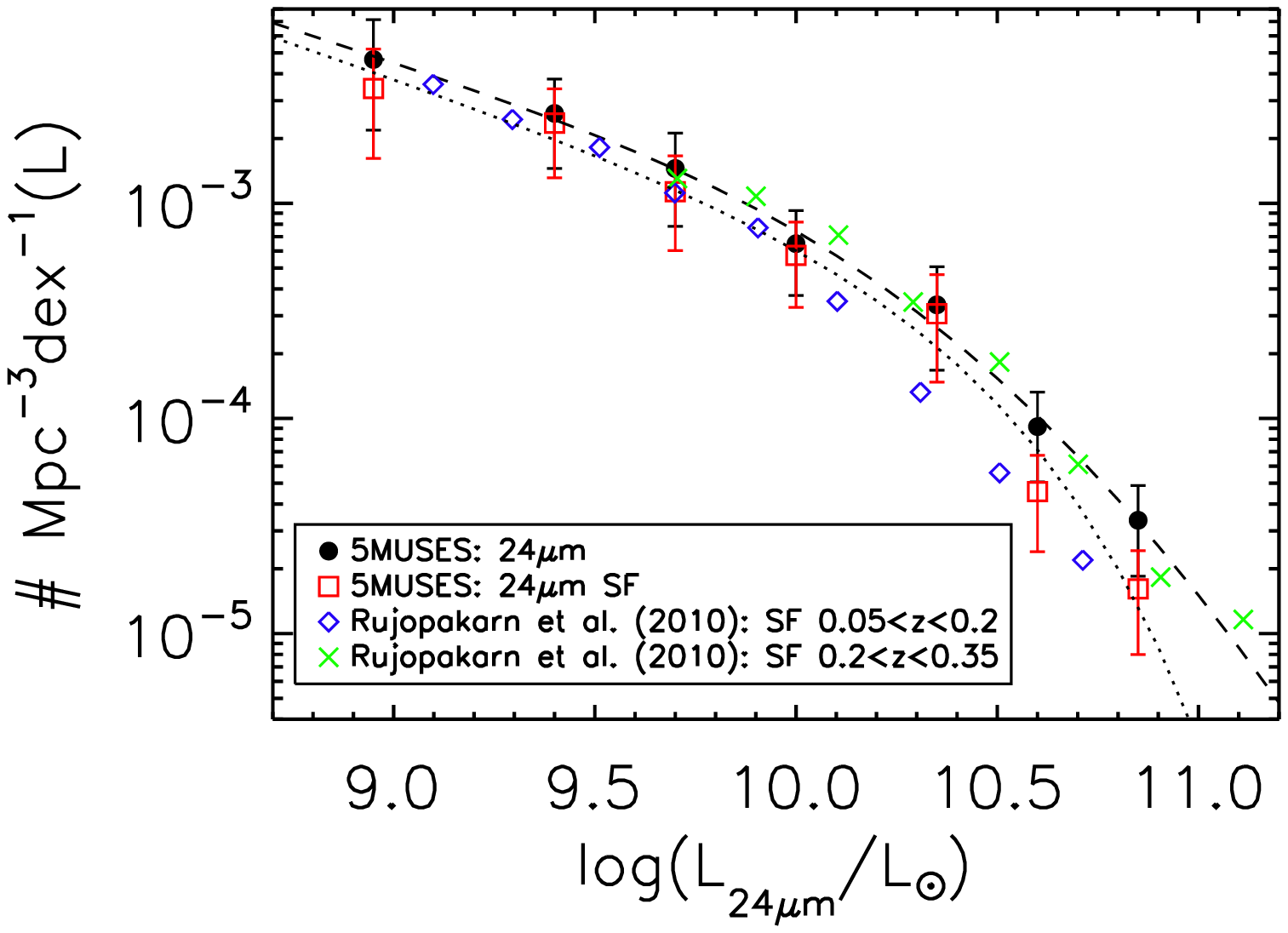}
  \caption{a) Left panel: A comparison of the 15\,$\mu$m luminosity
    function (black filled circles) and the 15\,$\mu$m star-formation
    luminosity function (red squares). The dashed line and the dotted
    line are the fits to the two LFs. The blue crosses represent the
    local AGN-excluded LF by \citet{Pozzi04} and appear to be in good
    agreement with our 15\,$\mu$m SF LF. The green diamonds represent
    the AGN LF derived from our sample and the dot-dashed line
    provides a fit to the AGN LF with a double power-law exponential
    function.b) Right panel: A comparison of the MIPS 24\,$\mu$m
    luminosity function (black filled circles) and star-formation
    luminosity function (red squares). The dashed lines are the fits
    to the total LFs with a double power law exponential function,
    while the dotted lines are the fits to the SF LFs with a Schechter
    function. The blue diamonds and green crosses represent the
    24\,$\mu$m SF LFs at 0.05$<$z$<$0.2 and 0.2$<$z$<$0.35 by
    \citet{Rujopakarn10}.}
  \label{fig:LF15_SF}
\end{figure}

{\bf Local 15\,$\mu$m SF LF: } Using the median estimate of the
star-formation fraction for each object, we derive the 15\,$\mu$m
star-formation luminosity by multiplying the 15\,$\mu$m monochromatic
continuum luminosity by the star-formation fraction at the
corresponding wavelength obtained from spectral decomposition. The
resulting individual SF luminosities are then used to build the SF
LF. On the left panel of Figure \ref{fig:LF15_SF}, we show the
15\,$\mu$m SF LF (red squares) at z$<$0.3. The black dotted line is a
fit to the LF adopting the Schechter function. 
\begin{equation}
\phi(L)=\frac{dN(L)}{dVdlog_{\mathrm{10}}(L)}=\phi^\star({\frac{L}{L^\star}})^{1-\alpha}exp{(-\frac{L}{L^\star})}
\end{equation}

We overplot the local 15\,$\mu$m LF for normal spiral and starburst
galaxies by \citet{Pozzi04} as the blue crosses. These authors
analyzed data from the ELAIS southern fields and excluded AGN in their
study. Their SFLF appear to be in good agreement with our results.  We
also overplot the total 15\,$\mu$m LF as the black filled circles on
the left panel of Figure \ref{fig:LF15_SF}, and the dashed line is a
fit to the total 15\,$\mu$m LF with double power-law exponential
function. The two LFs, the total and the star-formation LFs at
15\,$\mu$m, differ significantly at L$_{\rm 15\mu
  m}>10^{10}$L${_\odot}$. This is presumably due to the AGN
contribution at high luminosities. The AGN LFs display distinctly
different shapes from SF LF as has been shown by \citet{Matute06,
  Hopkins07}, etc. When the luminosity increases, the AGN contribution
also increases progressively. This AGN component reveals its presence
in the total 15\,$\mu$m LF by requiring a different slope in the fit
at high luminosity. In Figure \ref{fig:LF15_SF}, we have overplotted
the 15\,$\mu$m AGN LF as green diamonds and fit the data with a double
power-law exponential function. Clearly, the AGN LF presents a much
shallower slope at high luminosities. On the other hand, the
15\,$\mu$m SF LF drops quickly at the bright end and it could be fit
well with a Schechter function. This has already been seen in the
local universe for the 8\,$\mu$m LF of star forming galaxies by
\citet{Huang07}, as well as the 15\,$\mu$m AGN-corrected LF at
0.6$<$z$<$0.8 by \citet{Fu10}.

{\bf Local 24\,$\mu$m SF LF: } We repeat the same steps used to derive
the 15\,$\mu$m SF LF at 24\,$\mu$m and show the 24\,$\mu$m SF LF on
the right panel of Figure \ref{fig:LF15_SF}. Again, we observe the
departure of the total and SF LFs at 24\,$\mu$m at the bright end and
the best-fit parameters are reported in Table
\ref{LFpara}. \citet{Rujopakarn10} have identified AGN from optical
spectroscopy and derived 24\,$\mu$m SF LF in several redshift
bins. Their comparison of the 24\,$\mu$m total LF and the 24\,$\mu$m
SF LF shows a very similar trend as what we have observed from our
sample. We also overplot their 24\,$\mu$m SF LFs at 0.05$<$z$<$0.2 and
0.2$<$z$<$0.35 as the blue diamonds and green crosses on Figure
\ref{fig:LF15_SF}.

\subsection{Discussion}

As the main luminous phenomena in the universe, star formation and AGN
activities have been extensively studied across all
wavelengths. Despite the amount of effort to quantify the SF/AGN
contribution and explore its evolution with luminosity/redshift, no
fair comparison can be made unless truly equivalent samples are
studied. Our 5MUSES sample, after correcting for selection effects,
essentially defines a relatively bright IR selected unbiased sample,
which is critical for understanding the galaxy evolution and energy
balance in a cosmological context. In this subsection, we discuss how
the SF/AGN fraction varies with wavelengths, luminosity and redshift.

\subsubsection{SF/AGN fraction at mid- and total IR}

In Section 4, we have derived the 15\,$\mu$m luminosity function for
the entire sample at z$<$0.3, as well as the 15\,$\mu$m star-formation
luminosity function. Integrating the differential LF, we estimate the
luminosity density at 15\,$\mu$m to be
1.5$\pm$0.3$\times$10$^7$\,L$_\odot\,$Mpc$^{-3}$ to which the SF
contribution is 9.0$\pm$2.2$\times$10$^6$\,L$_\odot\,$Mpc$^{-3}$. This
gives a SF fraction of $\sim$58$\pm$19\% to the integrated 15\,$\mu$m
luminosity density. Then we extrapolate our results at 15\,$\mu$m to
the total IR.  Following the technique described in \citet{Wu10}, we
convert 15\,$\mu$m luminosity density to the total IR luminosity
density for the AGN component and the SF component
separately. Although the uncertainty in the SF fraction in total IR
will be significantly higher because the error in converting L$_{15\mu
  m}$ to L$_{\rm IR}$ also comes into play, this is still a critical
quantity to obtain, especially for studies of the distant universe,
where the PAH features are more difficult to measure, or the PAHs in
high-redshift galaxies might have different properties (e.g. larger
EWs) for similar dust mass fractions. We find that the star-formation
contribution comes up to 83\% of the total IR luminosity density. This
is understandable as the SED of a star-forming galaxies is much
steeper than AGN, thus the FIR emission will be dominated by SF. If we
convert the IR luminosity for the star-formation component to the
star-formation rate and integrate over cosmic time, we find our
derived star formation rate density are consistent with the dust
extinction corrected values of \citet{Hopkins06} (see also
\citet{Madau96, Lilly96}).  We repeat the same exercise at
24\,$\mu$m. Integrating the 24\,$\mu$m LF, we find the luminosity
density at 24\,$\mu$m to be
3.4$\pm$1.1$\times$10$^7$\,L$_\odot$\,Mpc$^{-3}$ while the SF
luminosity density is
2.6$\pm$1.1$\times$10$^{7}$\,L$_\odot$\,Mpc$^{-3}$. This gives a SF
fraction of $\sim$78$\pm$42\% at 24\,$\mu$m, higher than the number at
15\,$\mu$m, while consistent with the concept that SF becomes more
dominant at longer wavelength. We again convert the integrated
24\,$\mu$m luminosity to total IR luminosity density for star-forming
systems and AGN separately and find the star-formation fraction to
increase to $\sim$89\%, in agreement with the estimate from
15\,$\mu$m. If we take the average of the two (86\%), then the SF
fraction in the total IR luminosity density we have derived is higher
than the study of \citet{Veilleux09} for local ULIRGs, while more
consistent with \citet{Nardini08}. However, we need to bear in mind
that both Veilleux and Nardini study the ULIRG population, while the
luminosity of our sample is mostly in the range of
10$^{9.0}L_\odot$to 10$^{12.0}L_\odot$. Recent study of
\citet{Petric10} derived an average AGN fraction of 12\% in the total
IR for LIRGs, consistent with our estimate.

\begin{figure}
  \epsscale{1.2}
  \plottwo{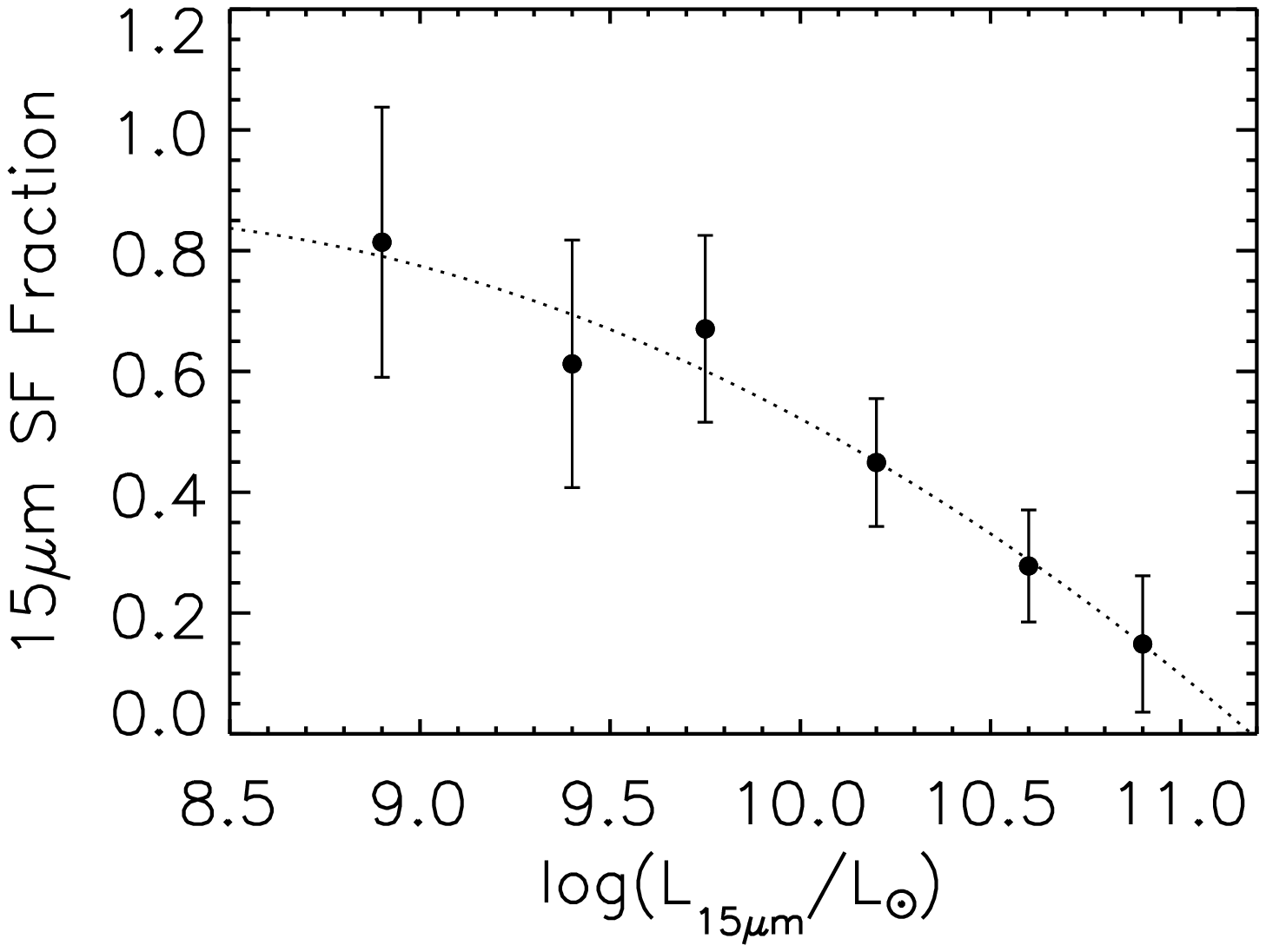}{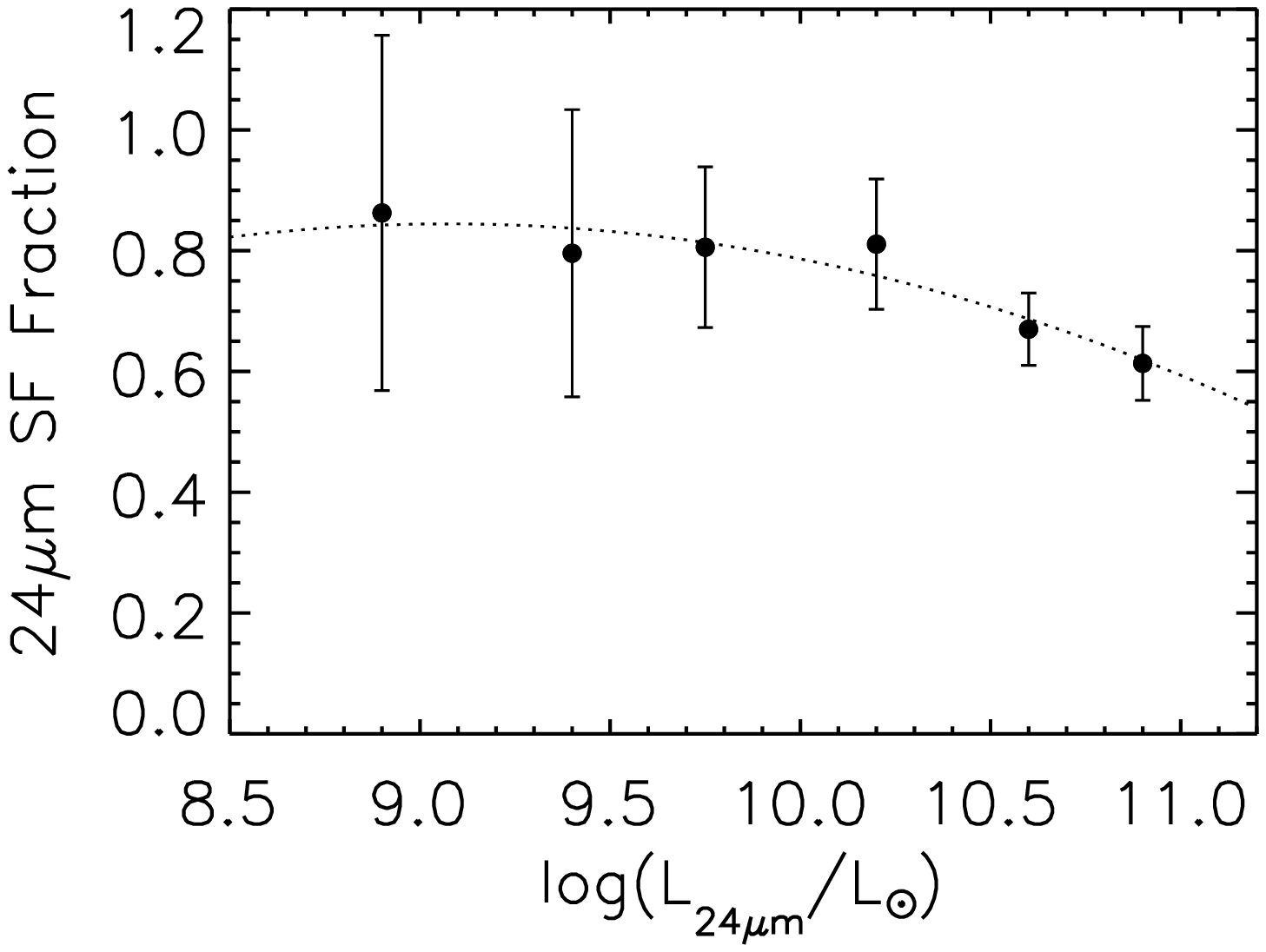}
  \caption{a) Left panel: The fractional contribution of the
    star-formation luminosity in each luminosity bin at
    15\,$\mu$m. The dotted line is a second order polynomial fit to
    the data. b) Same as a), but for 24\,$\mu$m. }
  \label{fig:SF_fraction}
\end{figure}

\begin{figure}
  \epsscale{1.2}
  \plottwo{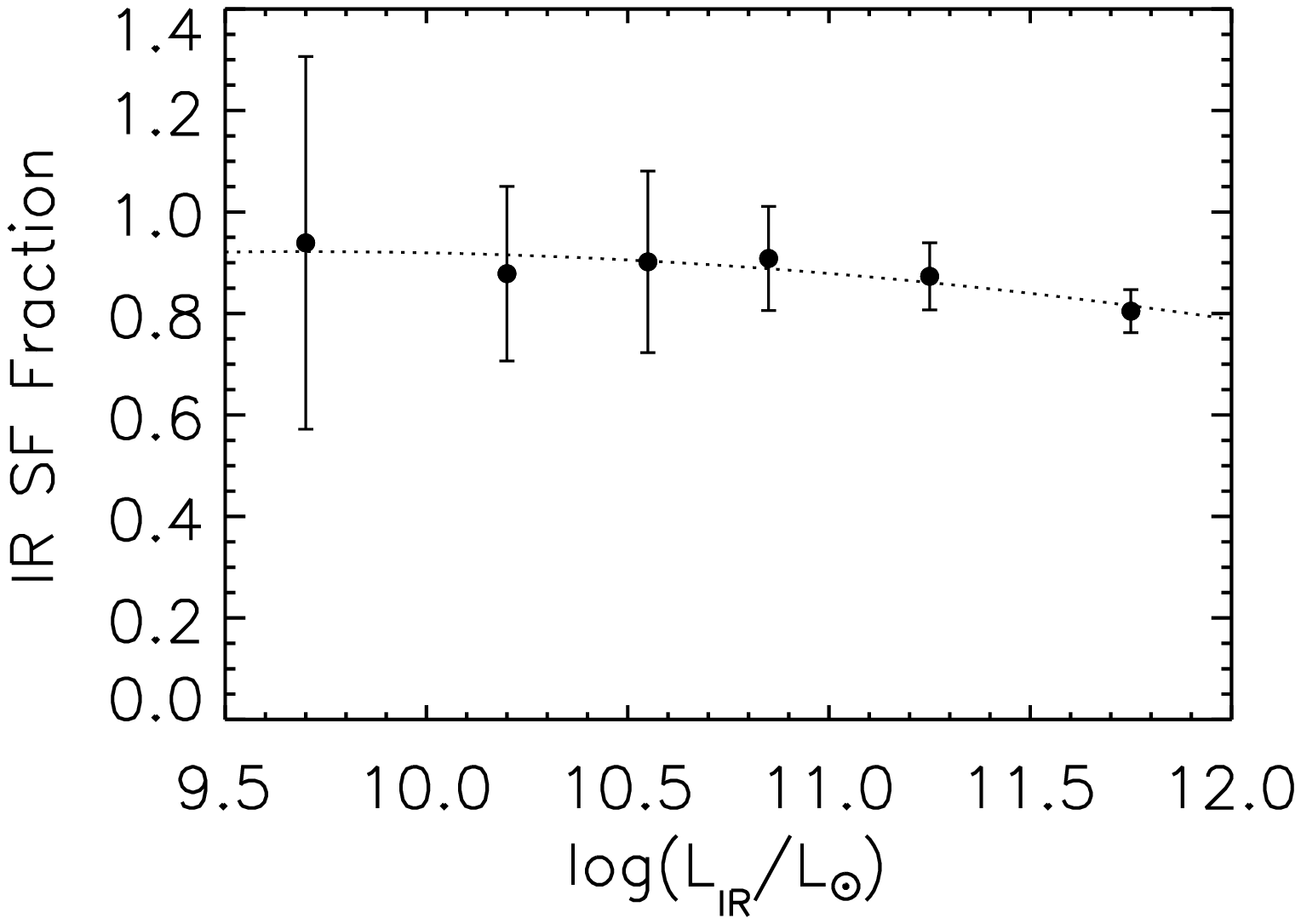}{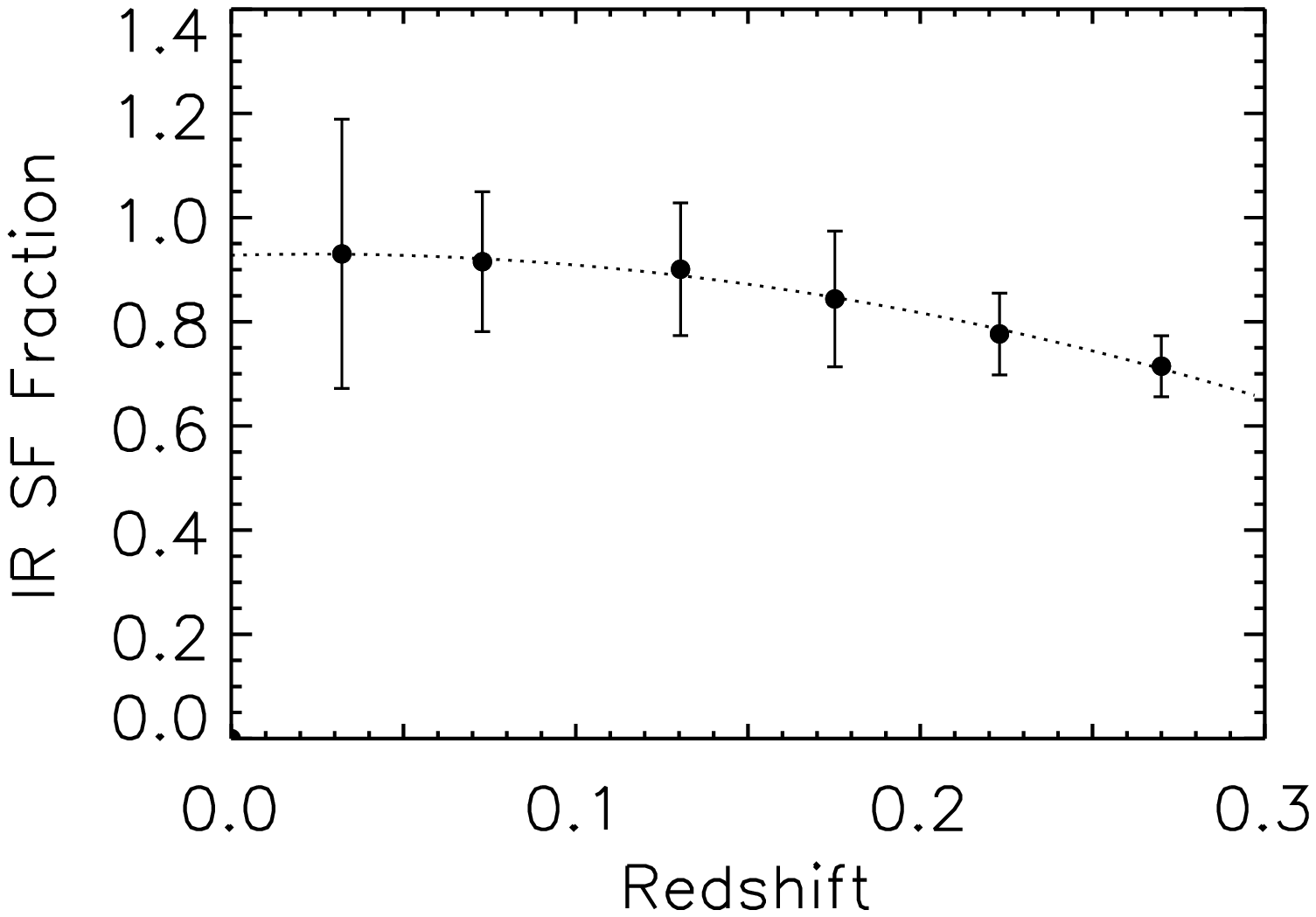}
  \caption{a) Left panel: The fractional contribution of the
    star-formation luminosity in each luminosity bin in the total IR
    luminosity, converted from the results at 24\,$\mu$m. The dotted
    line is a second order polynomial fit to the data. b) The SF
    fraction in the total IR luminosity versus the redshift. The
    dotted line is a second order polynomial fit to the data. }
  \label{fig:SF_fraction_LIR_z}
\end{figure}

As already noted (Figure \ref{fig:LF15_SF}), the 15\,$\mu$m SF LF
departs from the 15\,$\mu$m LF most significantly at the bright end,
and this is again observed at 24\,$\mu$m. It suggests that the SF
fraction is a function of luminosity \citep{Yuan10, Hopkins10}. In
order to quantify how this fraction varies, we plot on the left panel
of Figure \ref{fig:SF_fraction} the SF fraction at 15 \,$\mu$m versus
the 15\,$\mu$m luminosity. We divide our sources into several
luminosity bins and estimate the contribution of SF in each luminosity
bin. The error bar represents the Poisson noise in each bin. We find
that the SF fraction decreases as the 15\,$\mu$m luminosity
increases. This is expected since AGN plays a more important role in
the energy budget for more distant, thus more luminous sources in our
sample (see Figure \ref{fig:SF_fraction_LIR_z}). On the right panel of
Figure \ref{fig:SF_fraction}, we plot the SF fraction at 24\,$\mu$m
versus the 24\,$\mu$m luminosity. We again observe a trend of the SF
fraction decreasing with larger L$_{\rm 24\mu m}$, however, the
decline is much milder and the SF fraction is also higher at
24\,$\mu$m as compared to the 15\,$\mu$m.  Finally, we convert the
24\,$\mu$m SF luminosity and 24\,$\mu$m luminosity to L$_{\rm SFIR}$
and L$_{\rm IR}$, and show on the left panel of Figure
\ref{fig:SF_fraction_LIR_z}, how the total IR SF fraction varies as a
function of IR luminosity, while on the right panel, we display the SF
fraction versus redshift. We find that the SF fraction decreases with
redshift, while there is little dependence of SF fraction with L$_{\rm
  IR}$. The decrease of SF fraction with redshift can be understood
since our sample selects a higher fraction of AGN-dominated sources as
redshift increases (see also Figure \ref{fig:z_hist}). We observe a
mildly decreasing, or rather flat correlation between the IR SF
fraction and L$_{\rm IR}$ because: 1) At z$<$0.3, the luminosity of
our sources only ranges from 10$^{9}$L$_\odot$ to 10$^{12}$L$_\odot$
and the majority of the sources included in this study are SF
dominated; and 2) even for a source dominated by a powerful AGN in the
mid-IR, its FIR emission could still be powered by star-formation,
thus SF dominates the total IR luminosity. Because of the dominant
contribution of SF in the FIR luminosity, we do not observe a strong
dependence on L$_{\rm IR}$ for the SF fraction. This suggest that the
mid-IR might be a more reliable wavelength if one wants to study the
SF/AGN fraction.  The launch of the {\em Herschel} Space Telescope has
opened a new window for observing the cold dust in the universe. Data
from large area {\em Herschel} surveys, such as {\em HerMES}
\citep{Oliver10}, will provide essential constraints on the FIR
emission for sources in our sample, allowing a direct decomposition of
SF/AGN in the total IR luminosity. Future studies using 70-500\,$\mu$m
data from {\em Herschel} will help to further constrain the SF
fraction in the total IR luminosity density, and the associated
uncertainties.

\subsubsection{Comparison with LF at z=0.7}
\begin{figure}
  \epsscale{1.0}
  \plotone{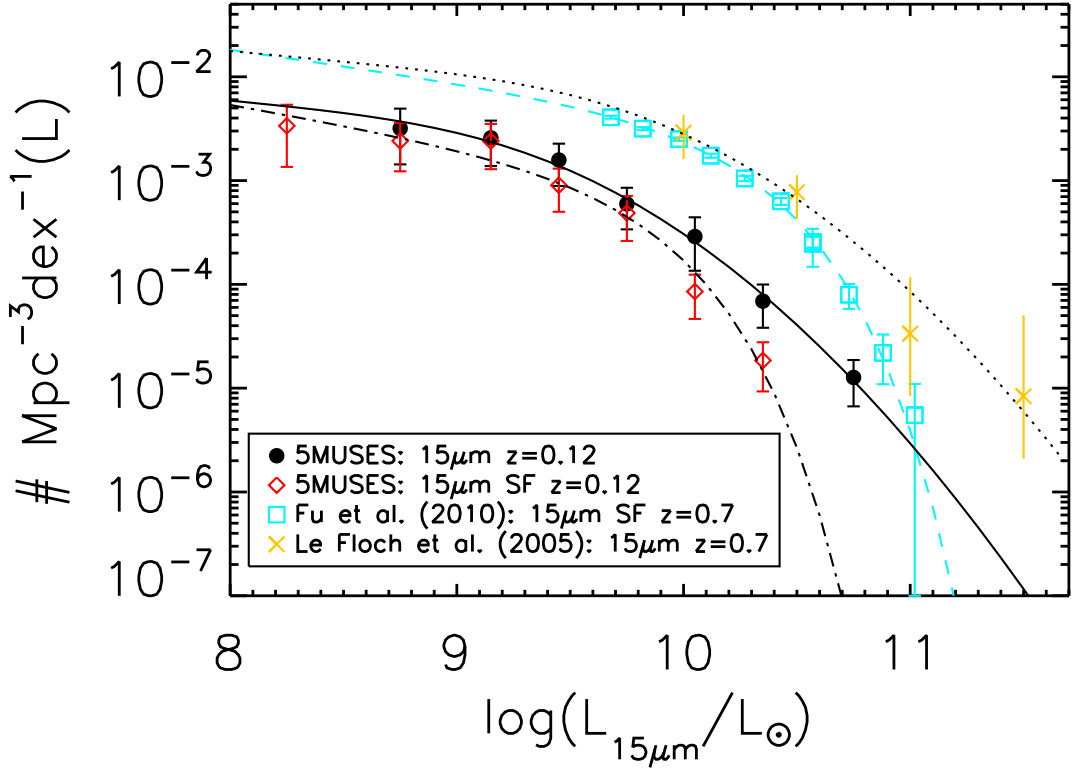}
  \caption{Comparison of LF at $<z>=0.12$ from our sample with LFs at
    $<z>=0.7$. The black filled circles represent the total 15\,$\mu$m
    LF and the red open diamonds represent the 15\,$\mu$m SF LF
    derived from our sample. The black dash-dotted line and the solid
    line are the best-fits to the SFLF and total LF derived from the
    5MUSES sample at 15\,$\mu$m. The blue open squares represent the
    15\,$\mu$m SF LF at z=0.7 from \citet{Fu10} and the yellow crosses
    represent the total 15\,$\mu$m LF from \citet{Lefloch05}. The blue
    dashed line is the best-fit Schechter function from \citet{Fu10}
    to the 15\,$\mu$m SF LF at z=0.7.  The black dotted line is
    derived by evolving the luminosity and density of LFs from 5MUSES
    by a factor of $\alpha_{\rm L}$=2.6 and $\alpha_{\rm D}$=2.1
    \citep{Lefloch05} and it appears to match well with the 15\,$\mu$m
    LF at z=0.7 by \citet{Lefloch05}.}
  \label{fig:LF_15_highz}
\end{figure}

In this section, we compare the LFs derived from the 5MUSES sample
with relevant work at higher redshift to understand the evolution
effects. The median redshift for our LF is 0.12. When compared with
15\,$\mu$m LF by \citet{Xu00} from ISO work, we find in general good
agreement between our LF and Xu's LF, while at the high luminosity
end, the two LFs show some discrepancy. Because of the higher median
redshift of the objects in the high luminosity bin of our LF, this
small discrepancy with the local LF hinted at evolution effects as a
function of redshift. We then compare our work with studies at high
redshift. 15 or 24\,$\mu$m LFs and LFs for star-forming galaxies have
been derived by several different groups \citep{Lefloch05, Magnelli09,
  Rujopakarn10, Fu10}. In addition to deriving the rest-frame
15\,$\mu$m LF from the 5MUSES sample, the availability of
5$-$35\,$\mu$m IRS spectra also allowed us to decompose the star
formation and AGN contributions to the IR spectral energy distribution
in a more precise way, and thus derive the SF LF using the SF
luminosity in each object from our sample. The most relevant work at
higher redshift is at z=0.7 by \citet{Fu10}. These authors use the IRS
spectra of a z=0.7 sample to estimate the SF and AGN contribution to
LF. In Figure \ref{fig:LF_15_highz}, we compare the 15\,$\mu$m SF LF
(red diamonds) from 5MUSES with the corresponding 15\,$\mu$m SF LF at
z=0.7 by \citet{Fu10} (blue squares). On the same figure, we also
overplot the total 15\,$\mu$m LF derived from 5MUSES sample (black
filled circles) and its counterpart at z=0.7 from the work of
\citet{Lefloch05} (yellow crosses). We observe strong evolution
effects in both the total and the SF LFs. With data only in two
redshift bins (z=0.12 and z=0.7), we were not able to place stringent
constraints on the amount of evolution on density ($\alpha_{\rm D}$)
and luminosity ($\alpha_{\rm L}$), however, if we adopt the values
proposed by \citet{Lefloch05} and evolve our total 15\,$\mu$m LF at
z=0.12 to z=0.7 by a factor of $\alpha_{\rm D}$=2.1 and $\alpha_{\rm
  L}$=2.6, we find a good match between the evolved LF (black dotted
line) and the observed data (yellow crosses).

\begin{figure}
  \epsscale{1.0}
  \plotone{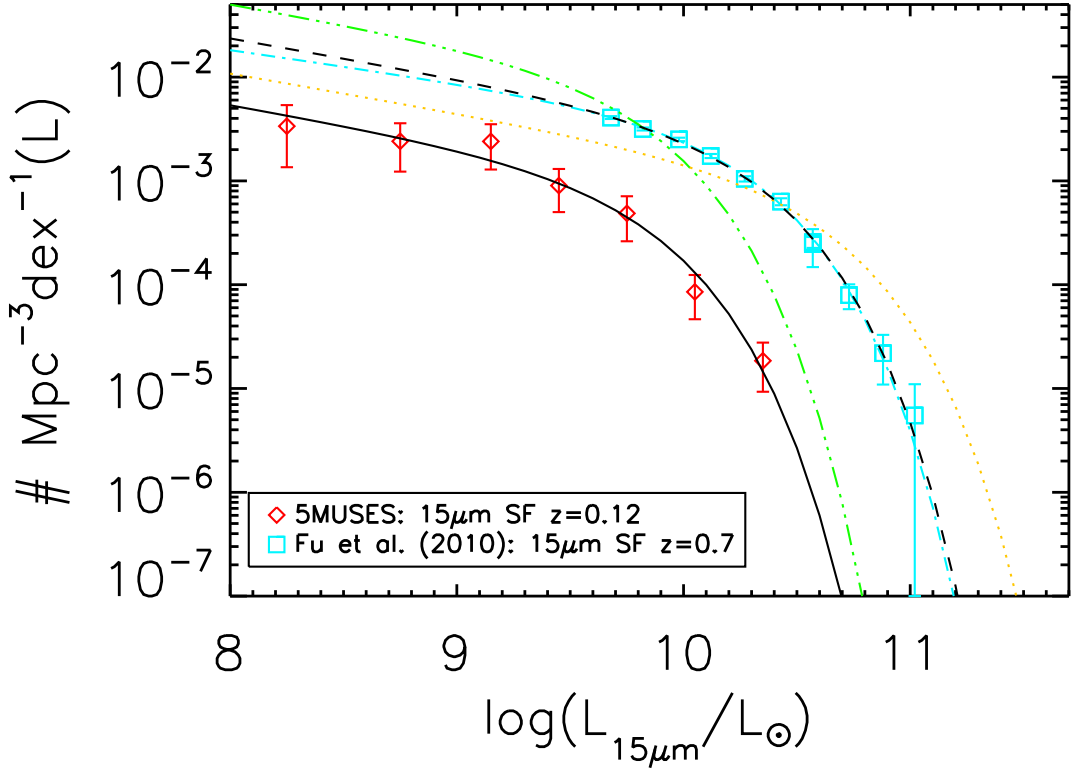}
  \caption{Comparison of the 15\,$\mu$m SF LF from 5MUSES (red
    diamonds) at $<z>=0.12$ and from \citet{Fu10} (blue squares) at
    $<z>=0.7$. The black solid line and the blue dash-dotted lines are the
    best fits to the two LFs respectively. The green dash-dot-dot line
    represent the best fit when only evolution on density is allowed
    ($\alpha_{\rm D}$=5.3). The yellow dotted line represents the best
    fit when only evolution on luminosity is allowed ($\alpha_{\rm
      L}$=4.3). The black dashed line represents the best fit when
    both density and luminosity evolution is allowed ($\alpha_{\rm
      D}$=2.5, $\alpha_{\rm L}$=2.6). }
  \label{fig:SFLF_evolve}
\end{figure}

We follow a similar approach to evolve the 15\,$\mu$m SFLF from z=0.12
to z=0.7. In Figure \ref{fig:SFLF_evolve}, we show that if we evolve
our 15\,$\mu$m SFL F at z=0.12 (black solid line) by a factor of
$\alpha_{\rm D}$=2.5 and $\alpha_{\rm L}$=2.6, it matches very well
(reduced $\chi^2=0.6$) with observed rest-frame 15\,$\mu$m SF LF at
z=0.7 from \citet{Fu10}'s work (blue dashed line). Could a pure
density evolution or luminosity evolution explain the difference we
observe in the SFLF at z=0.12 and z=0.7? In Figure
\ref{fig:SFLF_evolve}, we show the best fit to the SF LF at z=0.7 if
we only allow density evolution on our local SFLF (green dash-dot-dot
line). The best fit returns $\alpha_{\rm D}$=5.3 and the reduced
$\chi^2$=65.6. On the same figure, we also show the best fit when only
luminosity evolution is allowed (yellow dotted line). We find
$\alpha_{\rm L}$=4.3 and the reduced $\chi^2$=70.0. Even though we
only have data in two redshift bins, which makes it difficult to place
stringent constraints on the amount of evolution we require on
luminosity and density, we can at least confirm from our work that
neither pure luminosity nor pure density evolution is sufficient to explain the
difference between SF LFs at z=0.12 and z=0.7.  

We should however point out some caveats. Although both our work and
\citet{Fu10}'s work used IRS spectra as diagnostic tools for
distinguishing SF and AGN, our LF is derived by separating the energy
density contribution in each source, while they use IRS spectra to
determine the major source of energy in each object, and then remove
the AGN dominated sources when building their SFLF. Because they only
have IRS spectra for $\sim$ 40 galaxies from their flux limited
sample, the low end of their SFLF is obtained by shifting the MIPS
24\,$\mu$m luminosity to rest-frame 15\,$\mu$m (z=0.7), and they
assign these low luminosity objects as dominated by star
formation. Given our very rigorous analysis, we clearly see some
differences in the total LF and the SFLF at 15\,$\mu$m even at the low
luminosity end (see the comparison of the black dot-dashed line and
the black solid line), thus \citet{Fu10} might have overestimated the
SFLF below L$_{\rm 15\mu m}$=10$^{10.5}$L$_\odot$.

\section{Conclusions}
We have analyzed a Spitzer spectroscopic sample of 24\,$\mu$m selected
objects in the SWIRE and XFLS fields to derive the 15 and 24\,$\mu$m
luminosity functions at 0$<$z$<$0.3. When combined with local and
high-redshift studies, this provides critical information on
understanding the evolution of energy budgets over the past three
billion years (z$<$0.3). Our conclusions are summarized as follows:

1. We have derived 24 and 15\,$\mu$m LFs for our sample at
z$<$0.3. The availability of the 5-35\,$\mu$m allows us to make
K-corrections directly using the observed SED. The 24 and 15\,$\mu$m
LFs display rather shallow slopes at the bright end, which is due to the
increase of AGN contribution in more luminous systems.

2. Using the 5-35\,$\mu$m IRS spectra, we have decomposed the 5MUSES
objects into star-formation and AGN components. The SF fraction is
taken to be the median likelihood of the probability density function.
We calculate the 15 and 24\,$\mu$m SF luminosities in each object, and
subsequently build the SF LFs. The SF LFs can be described with a
Schechter function.

3. We have also estimated the SF contribution to the integrated 15 and
24\,$\mu$m luminosity density for our sample. The SF fraction is found
to be 58\% at 15\,$\mu$m and it goes up to 78\% at 24\,$\mu$m. Using
the conversion factor from L$_{\rm MIR}$ to L$_{\rm IR}$ for
star-forming galaxies and AGNs respectively, we found the SF fraction
to be $\sim$86\% in the total IR luminosity density.

4. The SF fraction is also found to be a function of
luminosity/redshift,decreasing as luminosity or redshift increase,
while the trend is more obvious in mid-IR, suggesting that mid-IR
wavelength is more sensitive to the presence of AGN.

5. Both luminosity and density evolution are required to explain the
difference in the observed SFLF between this sample and similar
studies at z$\sim$0.7.

\acknowledgments

We thank J.-S. Huang, V. Charmandaris, H. Fu and E. LeFloc'h for
insightful discussion. This work was based on observations made with
the {\em Spitzer Space Telescope}, which is operated by JPL/Caltech
under a contract with NASA. The observations are associated with the
{\em Spitzer} Legacy Program 40539. The authors acknowledge support by
NASA through awards issued by JPL/Caltech. This research has made use
of the NASA/IPAC Extragalactic Database (NED) which is operated by the
Jet Propulsion Laboratory, California Institute of Technology, under
contract with the National Aeronautics and Space Administration.

\begin{deluxetable}{cccccccc}
  \tabletypesize{\footnotesize} 
  \setlength{\tabcolsep}{0.002in}
  \tablecaption{mid-IR Luminosity Function Derived from the $1/V_{\rm
      max}$ Method\label{LFfit}} 
  \tablewidth{0pc} 
  \tablehead{
    \colhead{logL$_{\rm 24\mu m}$} & \colhead{$\phi$} & \colhead{logL$_{\rm 15\mu m}$} & \colhead{$\phi$} & \colhead{logL$_{\rm 24\mu mSF}$} & \colhead{$\phi$} & \colhead{logL$_{\rm 15\mu mSF}$} & \colhead{$\phi$}\\ 
    \colhead{(L$_\odot$)} & \colhead{[Mpc$^{-3}$dex$^{-1}(L)$]} & \colhead{(L$_\odot$)} & \colhead{[Mpc$^{-3}$dex$^{-1}(L)$]} & \colhead{(L$_\odot$)} & \colhead{[Mpc$^{-3}$dex$^{-1}(L)$]} & \colhead{(L$_\odot$)} & \colhead{[Mpc$^{-3}$dex$^{-1}(L)$]}} 
    \startdata 

    8.95 & 4.66$\pm$2.47$\times$10$^{-3}$ & 8.75 &  3.18$\pm$1.75$\times$10$^{-3}$ & 8.95 & 3.41$\pm$1.80$\times$10$^{-3}$ & 8.25 & 3.36$\pm$2.01$\times$10$^{-3}$ \\
    9.40 & 2.62$\pm$1.16$\times$10$^{-3}$ & 9.15 &  2.58$\pm$1.12$\times$10$^{-3}$ & 9.40 & 2.36$\pm$1.05$\times$10$^{-3}$ & 8.75 & 2.41$\pm$1.18$\times$10$^{-3}$ \\ 
    9.70 & 1.46$\pm$0.67$\times$10$^{-3}$ & 9.45 &  1.58$\pm$0.69$\times$10$^{-3}$ & 9.70 & 1.13$\pm$0.53$\times$10$^{-3}$ & 9.15 & 2.40$\pm$1.11$\times$10$^{-3}$ \\ 
    10.00& 6.49$\pm$2.76$\times$10$^{-4}$ & 9.75 & 5.96$\pm$2.57$\times$10$^{-4}$ & 10.00 & 5.73$\pm$2.45$\times$10$^{-4}$ & 9.45 & 9.01$\pm$4.01$\times$10$^{-4}$ \\ 
    10.35& 3.37$\pm$1.70$\times$10$^{-4}$ & 10.05 & 2.89$\pm$1.54$\times$10$^{-4}$ & 10.35 & 3.06$\pm$1.59$\times$10$^{-4}$ & 9.75 & 4.87$\pm$2.25$\times$10$^{-4}$ \\ 
    10.60& 9.16$\pm$4.09$\times$10$^{-5}$ & 10.35 & 6.95$\pm$3.07$\times$10$^{-5}$ & 10.60 & 4.58$\pm$2.17$\times$10$^{-5}$ & 10.05 & 8.54$\pm$3.88$\times$10$^{-5}$ \\ 
    10.85& 3.40$\pm$1.51$\times$10$^{-5}$ & 10.75 & 1.36$\pm$0.60$\times$10$^{-5}$ & 10.85 & 1.64$\pm$0.81$\times$10$^{-5}$ & 10.35 & 1.87$\pm$0.92$\times$10$^{-5}$ \\ 
  \enddata
\end{deluxetable}

\begin{deluxetable}{cccccc}
  \setlength{\tabcolsep}{0.1in}
  \tablecaption{Results of the FITs to the mid-IR Luminosity functions\label{LFpara}}
  \tablewidth{0pc}
  \tablehead{
  \colhead{} & \colhead{$\alpha$} & \colhead{$\sigma$} & \colhead{logL$^\star$(L$_\odot$)} & \colhead{log$\phi$[Mpc$^{-3}$dex$^{-1}(L)$]} & \colhead{$\rho$(L$_\odot$Mpc$^{-3}$)}

  }
  \startdata
  LF (24\,$\mu$m)    & 1.61$\pm$0.62 & 0.57$\pm$0.30 & 9.63$\pm$1.42 & -2.72$\pm$0.98 & 3.36$\pm$1.06$\times$10$^7$ \\
  LF (15\,$\mu$m)    & 1.20 (fixed) & 0.65$\pm$0.22 & 8.71$\pm$1.35 & -2.24$\pm$0.99 & 1.54$\pm$0.32$\times$10$^7$ \\
  LF (24\,$\mu$m SF) & 1.63$\pm$0.28 &\nodata & 10.37$\pm$0.28 & -3.27$\pm$0.40 & 2.61$\pm$1.14$\times$10$^7$ \\
  LF (15\,$\mu$m SF) & 1.38$\pm$0.20 &\nodata &  9.76$\pm$0.16 & -2.93$\pm$0.26 & 8.99$\pm$2.20$\times$10$^6$ \\
  \enddata
\end{deluxetable}

\end{document}